\documentclass[12pt]{paper}
\usepackage[T1]{fontenc}
\usepackage[utf8]{inputenc}
\usepackage[english]{babel}
\usepackage[a4paper,top=3.5 cm,bottom=2.5 cm,left=2cm,right=2cm,heightrounded,bindingoffset=5mm]{geometry}
\usepackage{authblk}
\usepackage[authoryear]{natbib}
\usepackage{amsmath}
\usepackage{amsfonts}
\usepackage{amssymb}
\usepackage{graphicx}
\usepackage{latexsym}
\usepackage{array}
\usepackage{booktabs}
\usepackage{float}
\usepackage{multirow}
\usepackage{subfig}
\usepackage{color}
\usepackage[dvipsnames]{xcolor}
\usepackage{algorithm} 
\usepackage{soul}
\usepackage{rotating}
\floatname{algorithm}{Algorithm}

\usepackage{setspace}
\usepackage{marvosym}

\date{}

\begin{document}

\author[1]{Fabio Divino 
	\footnote{\Letter \; \textit{fabio.divino@unimol.it}}
	}
\author[2]{Johanna \"{A}rje}
\author[2]{Antti Penttinen}
\author[3]{Kristian Meissner}
\author[2]{Salme K\"{a}rkk\"{a}inen}
\affil[1]{\textit{Division of Physics, Computer Science and Mathematics - Department of Biosciences, University of Molise}}
\affil[2]{\textit{Department of Mathematics and Statistics, University of Jyv\"{a}skyl\"{a}}}
\affil[3]{\textit{Programme for Environmental Information, Finnish Environment Institute SYKE, Jyv\"{a}skyl\"{a}}}
\title{Empirical Bayes to assess ecological diversity and similarity with overdispersion in multivariate counts}
\maketitle
\begin{abstract}
The assessment of diversity and similarity is relevant in monitoring the status of ecosystems. The respective indicators are based on the taxonomic composition of biological communities of interest, currently estimated through the proportions computed from sampling multivariate counts. In this work we present a novel method able to work with only one sample to estimate the taxonomic composition when the data are affected by overdispersion. The presence of overdispersion in taxonomic counts may be the result of significant environmental factors which are often unobservable but influence communities. Following the empirical Bayes approach, we combine a Bayesian model with the marginal likelihood method to jointly estimate the taxonomic proportions and the level of overdispersion from one sample of multivariate counts. Our proposal is compared to the classical maximum likelihood method in an extensive simulation study with different realistic scenarios. An application to real data from aquatic biomonitoring is also presented. In both the simulation study and the real data application, we consider communities characterized by a large number of taxonomic categories, such as aquatic macroinvertebrates or bacteria which are often overdispersed. The applicative results demonstrate an overall superiority of the empirical Bayes method in almost all examined cases, for both assessments of diversity and similarity. We would recommend practitioners in biomonitoring to use the proposed approach in addition to the traditional procedures. The empirical Bayes estimation allows to better control the error propagation due to the presence of overdispersion in biological data, with a more efficient managerial decision making.
\end{abstract}
\section{Introduction}\label{intro}
Statistical measures of diversity and similarity \citep{rao:1980} are fundamental indicators for the assessment of biodiversity and the monitoring of biodiversity dynamics in space and time \citep{pielou:1975, magurr:2004, gotcha:2013}. This is relevant in environmental management as changes in observed trends and patterns are often related to changes in the status of ecosystems \citep{spelle:2005, matwhi:2015}. To assess such changes, a large variety of indices are typically used \citep{birkal:2012}.\\ 
In general, statistical measures of diversity and similarity are based on the sampling estimates of proportions which describe the composition of biological communities in categories of taxonomic rank (e.g. species or genera). Commonly, these estimates are the sampling proportions obtained through the maximum likelihood method under the \textit{Multinomial} model \citep{pielou:1969}.\\ 
Despite this assumption, estimates of taxonomic composition could present unexpected variability across the environmental domain where the data are collected. For instance, in aquatic biomonitoring, benthic taxa are an important indicator to assess and control the quality of rivers, lakes, and lagoons \citep{pfebau:1998, birkal:2012}. However, estimates of species proportions from distinct samples of the same aquatic environment often exhibit higher variation than what is expected by the adopted model \citep{wargut:2011, qianal:2014}. This lack of homogeneity in the sampling variability may be due to the action of significant factors which are unobservable but influence the behaviour of the individuals in the population, introducing extra variation in the data, with bias and noise in the inferential results.\\ 
The described phenomenon is well known as overdispersion \citep{clapha:1936}, formalized with count data in the seminal paper by \cite{blifis:1953} and then extended in \cite{mccnel:1983}, \cite{cox:1983}, and \cite{breslow:1984}. Overdispersion represents extra variation of the response variable with respect to the hypothesized model and \textit{``can be interpreted as brought about by failures of some basic assumption of the model''} \citep{xekala:2014}. In particular, with ecological count data, overdispersion may be observed because the adopted model does not account for extra variation caused by the effect of clustering among the individuals and the presence of heterogeneity in the composition of the population \citep{richar:2008, etteral:2009, linman:2011, harris:2014}.\\ 
In the literature there are several approaches to account for the presence of overdispersion with count data and they can be combined with each other. For instance, the use of appropriate designs, such as stratified and adaptive sampling \citep{thomps:1992,levlem:2008,mannav:2015}; the introduction of specific overdispersion parameters in the statistical models \citep{mccnel:1983,lindse:1995}; the application of regression methods when information about the sources of variability is available \citep{mccnel:1983,lindse:1995,hilbe:2011,linman:2011}; or the adoption of suitable distributions, such as the zero-inflated \textit{Poisson}, the \textit{Negative Binomial}, or the \textit{Negative Multinomial} distributions \citep{etteral:2009,kokone:2014}.\\
Like other sources of variability, the added variation introduced by overdispersion can result in less accurate estimates of the indices used. In the context of ecosystem management, the consequences of the ensuing error propagation into ecological status assessment of ecosystems can cause costly erroneous managerial decisions that could result in either unnecessary restoration efforts or in the failure to conduct restoration efforts where they are in fact needed.\\
In the present paper we propose a Bayesian model to estimate the composition of biological populations in taxonomic categories when the observed multivariate counts are affected by overdispersion. In the construction of our proposal, we consider the compound method that was introduced by \cite{greyul:1920} in the field of actuarial statistics; further specified in \cite{feller:1943}, \cite{batney:1952}, \cite{patil:1962,patil:1965}, and \cite{mosima:1962}; with applications presented in \cite{nelson:1984,nelson:1985}, \cite{chenli:2013}, and \cite{valpha:2013}. The model is based on the conjugate property of the \textit{Dirichlet} distribution for the \textit{Multinomial} likelihood \citep{diaylv:1979}. The inferential approach is empirical Bayes \citep{carlou:2000} and the concentration parameter of the adopted (noninformative) symmetric \textit{Dirichlet} prior is estimated by the maximum marginal likelihood which is applied to the respective compound \textit{Dirichlet-Multinomial} model \citep{mosima:1962}.\\
As in \cite{nelson:1984,nelson:1985}, we do not consider the use of explicative variables. Our primary goal is a proper estimation of the taxonomic composition in order to achieve a more robust assessment of the biological diversity and similarity when data are affected by overdispersion and no covariates are available. To do so, we present a simple method able to work when only one sample of multivariate counts is observed. Our proposal is particularly appropriate to monitoring of diversity and similarity when the composition of the community is characterized by a large number of taxonomic categories with mostly few specimens, such as e.g. diatoms, aquatic macroinvertebrates, or bacteria.\\
To assess the impact of our proposed approach on the behaviour of indices, we present an extensive simulation study with several levels of overdispersion and sample size. We consider three theoretical types of communities: \textit{quasi uniform evenness}, with species distributed approximately with the same level of abundance; \textit{smooth evenness}, with a gradually increasing pattern of abundance; and \textit{concentrated evenness}, with significant abundance only in few of the species. Through the simulations, the empirical Bayes estimates are compared to the \textit{Multinomial} maximum likelihood estimates in terms of statistical results referred to a selected set of diversity and similarity indices. The chosen indicators in this paper are the Shannon entropy \citep{shawea:1963}, the Simpson diversity \citep{simpso:1949}, the  percent model affinity \citep{renkon:1938, novbod:1992}, and the Euclidean similarity \citep{cliste:1975}.\\
The paper is organized as follows. Section \ref{over} reviews the necessary background about overdispersion in multivariate counts and introduces our approach. In Section \ref{ebayes} we present the Bayesian model in detail and the empirical Bayes inference to estimate the composition proportions from one sample of data. In Section \ref{simul} we illustrate the results of the simulation study with application to the assessment of diversity and similarity whilst an application with data from aquatic biomonitoring is presented in Section \ref{appl}. Finally, in Section \ref{final} conclusions are drawn and potential extensions are discussed.
\section{Multivariate counts with overdispersion}\label{over}
\begin{figure}[H]
	\centering
	\includegraphics[scale=0.5]{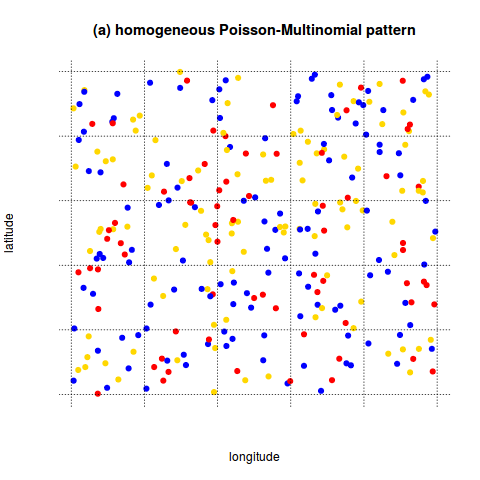}\includegraphics[scale=0.5]{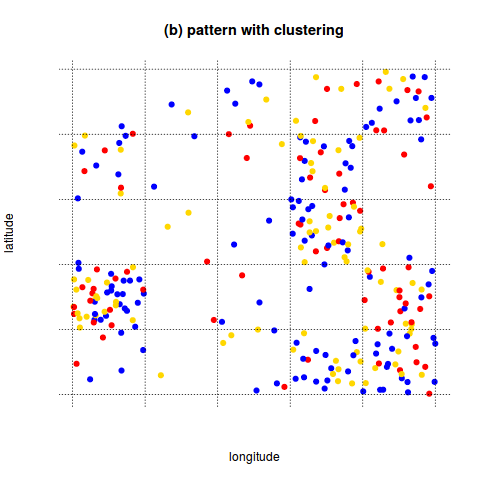}
	\includegraphics[scale=0.5]{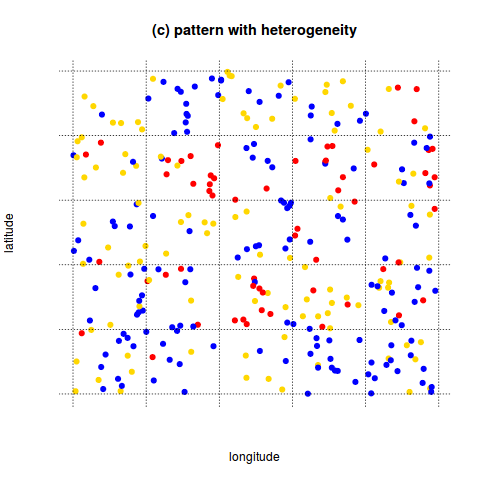}\includegraphics[scale=0.5]{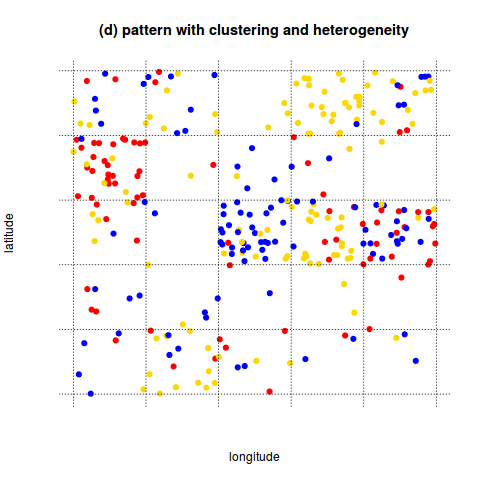}
	\caption{Simulated point patterns: (a) homogeneous pattern with the only Poisson-Multinomial variation, (b) pattern with overdispersion due to clustering among individuals, (c) pattern with overdispersion due to heterogeneity in composition, and (d) pattern with overdispersion due to clustering and heterogeneity. }\label{fig:exp}
\end{figure}

Let us consider a biological population of individuals $\mathcal{P}$ defined on a universe of interest $\mathcal{U}$ which is characterized by two parameters: a positive quantity $\lambda$, that represents the expected number of individuals in a specified universe unit $U$, and a vector $\underline{\pi}=(\pi_1,...,\pi_k)$ of nonnegative proportions normalized to unity, that represents the composition of the population $\mathcal{P}$ in a set of given $k$ categories $\{c_1,...,c_k \}$. In this setting, we partially follow terminology and notation as presented in \cite{steove:1994} and \cite{wangal:2012}, and refer to the universe $\mathcal{U}$ as the ecological environment of interest where the biological population $\mathcal{P}$ is defined on. Therefore, $\mathcal{P}$ represents an attribute or response of $\mathcal{U}$ \citep{steove:1994}.\\ 
In particular, we consider the case in which the environment $\mathcal{U}$ is a continuous physical space (such as the area of a forest or the benthic area of a lake, for instance), the population $\mathcal{P}$ is the set of individuals of interest living in the environment $\mathcal{U}$ (such as the birds in the forest or the benthic macroinvertebrates who populate the lake), the categories $\{c_1,...,c_k \}$ are taxonomic categories (such as the species of birds or the macroinvertebrate taxa), and the universe unit $U$ is defined as the sampling unit, that is the frame of environment specified in the sampling design that can be collected from $\mathcal{U}$ (such as the plot of fixed size in a grid overlapping the geographical area of the forest or the typical volume of water that can be sampled at any location of the area of the lake).\\
From a sampling point of view, $\lambda$ can be interpreted as the number of individuals expected in a potential random sample $S$ of unit size, that is the expected sampling abundance; and each proportion $\pi_j$ as the probability that an individual, randomly selected from the individuals collected in $S$, belongs to the respective taxonomic category $c_j$. Notice that, in this setting, the quantity $\lambda \pi_j$ is the expected total of the taxon $c_j$ and represents the tendency to observe individuals from that taxonomic category in a random sample of unit size. When a generic sample $S$ has a size different from the universe unit, the number of expected individuals is given by $\lambda v_S$, where $v_S$ is the relative size of $S$ with respect to the size of $U$. Without loss of generality, we consider samples of unit size under a simple random design \citep{thomps:1992,levlem:2008}.\\ 
In general, given a random sample $S$ of unit size, for each $j$ (with $j=1,...,k$), we denote by $x_{j}$ the count of individuals observed in the taxon $c_j$ and assume that each $x_{j}$ follows a conditionally independent \textit{Poisson} distribution with mean $\lambda \pi_j$
\begin{equation}\label{pois1}
p(x_{j}\mid \lambda, \pi_j)= \dfrac{(\lambda \pi_j)^{x_{j}}e^{-\lambda \pi_j}}{x_{j}!}.
\end{equation}
Hence, the total number of individuals $n = \sum_{j = 1}^k x_{j}$ observed in the sample $S$ is also \textit{Poisson} distributed with mean $\lambda$ 
\begin{equation}\label{pois2}
p(n \mid \lambda)= \dfrac{\lambda ^{n}e^{-\lambda}}{n!}.
\end{equation}
We refer to $n$ as the sample size in terms of the number of individuals collected in the sample $S$. Further, the conditional joint distribution of $\underline{x}=(x_{1},...,x_{k})$, given the vector $\underline{\pi}$ and the sample size $n$, is \textit{Multinomial}
\begin{equation}\label{mult}
p(\underline{x} \mid \underline{\pi},n) = n! \prod_{j=1}^{k} \dfrac{\pi_j^{x_{j}}}{x_{j}!} I (\Sigma_j x_j-n),
\end{equation}
where $I(\Sigma_j x_j-n)$, with $I(z)=1$ only when $z=0$ and null otherwise, is the indicator function to account for the constraint $\sum_{j=1}^k x_j=n$.
Commonly, a suitable estimate $\hat{\underline{\pi}}=(\hat{\pi}_1,...,\hat{\pi}_k)$ of the composition parameter $\underline{\pi}$ can be obtained by the maximum likelihood method with 
\begin{equation}\label{mle}
\hat{\pi}_j= \dfrac{x_j}{n},
\end{equation}
for every $j=1,...,k$ \citep{pielou:1969}.\\
A typical pattern of data described by combining Equation \eqref{pois2} with Equation \eqref{mult} is visualized in Figure \ref{fig:exp}.a in which a population of points is simulated over a spatial environment with the same level of abundance and with the same composition in three taxa (coloured in blue, red, and yellow respectively) across the whole area. The $5 \times 5$ grid of plots overlapping the environment denotes the set of potential samples of unit size that can be collected in $\mathcal{U}$. In this example, the expected number of points in each plot is $\lambda=12$ and the distribution into the three taxonomic categories is given by $\underline{\pi}=(0.40\;0.25\;0.35)$.\\
The pattern in Figure \ref{fig:exp}.a represents an ideal situation in which the population $\mathcal{P}$ responds to the environment $\mathcal{U}$ coherently with respect to the probabilistic model specified by Equations \eqref{pois2} and \eqref{mult}. The point pattern and its composition are homogeneous across the whole environmental space and this is the tacit general assumption underlying current bioassessments. Unfortunately, real situations may present patterns significantly different from such ideal configuration and it is not always possible to control for the potential lack of homogeneity, because the environmental determinants affecting the population $\mathcal{P}$ may be latent and unobservable.\\ 
To describe more realistic processes, it is necessary to interpret the parameters $\lambda$ and $\underline{\pi}$ as random quantities varying across the environmental domain $\mathcal{U}$. This approach allows us to include additional uncertainty in the sampling abundance and composition not present in the ideal pattern of Figure \ref{fig:exp}.a. In particular, the interpretation of $\lambda$ and $\underline{\pi}$ as random parameters allows us to describe the specific patterns of clustering in the population $\mathcal{P}$ and heterogeneity in its composition respectively. Due to the action of latent factors that can affect the behaviour of the individuals living in the environment $\mathcal{U}$, both heterogeneity and clustering are central sources of overdispersion \citep{xekala:2014}.\\
\begin{center}
<Figure \ref{fig:exp}>
\end{center}
In general, clustering occurs when the individuals of the population $\mathcal{P}$ do not occupy the space of the environment with a constant level of abundance \citep{xekala:2014}. Therefore, the parameter $\lambda$ changes over $\mathcal{U}$ and the concentration of individuals in each environmental unit may be different from each other. As a consequence, the data may present larger variation in the sample size than what is expected by the \textit{Poisson} model of Equation \eqref{pois2}. This experimental situation is visualized in Figure \ref{fig:exp}.b where the points in each plot are characterized by a specific value of expected abundance, with an evident clustering effect across the environment $\mathcal{U}$.\\ 
Heterogeneity occurs when the individuals of the population $\mathcal{P}$ occupy the space with the same level of concentration but with the taxonomic composition varying over $\mathcal{U}$ \citep{xekala:2014}. As a direct effect, the data may present extra variation in the sampling counts with different patterns of points across the environment not expected by the \textit{Multinomial} model in Equation \eqref{mult}. This is visualized in Figure \ref{fig:exp}.c where the compositions in three categories are different by plot, but no clustering effect in the population is present.\\
The most serious but very common situation is when latent factors induce both clustering in the population $\mathcal{P}$ and heterogeneity in its composition across the environment $\mathcal{U}$, with data that may present extra variation in both the sampling size and the taxonomic counts. A typical configuration of this type of samples is shown in Figure \ref{fig:exp}.d where the simulated points exhibit patterns with clusters of points and different compositions across the set of plots overlapping $\mathcal{U}$.\\
Hence, when ecologists use the empirical proportions in Equation \eqref{mle} to assess diversity and similarity, they are assuming the \textit{Poisson-Multinomial} hypotheses of Equations \eqref{pois2} and \eqref{mult} as valid. These hypotheses are underlying patterns like the one in Figure \ref{fig:exp}.a, but in most of real situations data exhibit patterns more similar to Figure \ref{fig:exp}.d. Therefore, the maximum likelihood estimation is inefficient in assessing diversity and similarity in communities where the combined effect of clustering and heterogeneity can produce configurations with high levels of overdispersion. It is necessary to consider more efficient methods when data show this type of patterns. In the following Section \ref{ebayes} we present in detail the empirical Bayes approach with the ability to overcome the aforementioned problems.
\section{Bayesian model and empirical Bayes estimation}\label{ebayes}
In general, when the parameters of interest are considered as random quantities, a direct approach to make inference about those parameters is the application of a Bayesian scheme. Therefore, the randomness necessary to account for extra variation in the data can be interpreted as prior uncertainty.\\ 
In the present setting, our goal is the estimation of the composition parameter $\underline{\pi}$, with respect to a population of individuals classified into $k$ pre-determined taxonomic categories. To account for the potential presence of overdispersion in data, we adopt the empirical Bayes approach \citep{carlou:2000}, with a Bayesian model based on the conjugate property of the \textit{Dirichlet} distribution for the \textit{Multinomial} likelihood \citep{diaylv:1979}. We assume that the parameter $\underline{\pi}$ follows a noninformative symmetric \textit{Dirichlet} prior, whose concentration parameter is estimated through the marginal likelihood obtained by the compound \textit{Dirichlet-Multinomial} model \citep{mosima:1962}. 
The marginalization of the likelihood combined with the prior represents a standard in empirical Bayes to get a prior guess of the hyperparameters from the data.\\ 
Notice that our proposal is designed for situations where only one sample is observed, with the restriction that the expected sample size $\lambda$ is considered unknown but not random. Therefore, in this setting, the specific overdispersion due to clustering cannot be properly modelled and can be only partially accounted for.\bigskip\\
\textit{Likelihood.} We consider that data are observed from one sample $S$ of unit size which is randomly collected from the environment $\mathcal{U}$. In particular, given the composition parameter $\underline{\pi}$ and the sampling abundance $n$, we assume that the sampling count data $\underline{x}$ follow the \textit{Multinomial} model in Equation \eqref{mult}.\bigskip\\
\textit{Prior.} We formalize the uncertainty about the composition parameter by a noninformative prior and assume that $\underline{\pi}$ is distributed following a symmetric \textit{Dirichlet} distribution
\begin{equation}\label{diric1}
p(\underline{\pi} ; \eta) = \Gamma (k\eta) \prod_{j=1}^{k} \dfrac{\pi_{j}^{\eta-1}}{\Gamma(\eta)},
\end{equation}
where $\Gamma(z)$ is the canonical gamma function and $\eta>0$ is the concentration parameter governing the level of prior uncertainty. In particular, when $\eta$ is small, the prior uncertainty is large while with large values of $\eta$ the prior uncertainty declines. Furthermore, notice that the distribution in Equation \eqref{diric1} is defined on the $k-1$ simplex $\Delta_{k-1}$, due to the constraint $\sum_{j=1}^k \pi_j=1$, with uniform expectation $E[\underline{\pi}]=(\frac{1}{k}, ... , \frac{1}{k})$ over the taxonomic categories. \bigskip\\
\textit{Posterior.} Through the Bayes theorem, the posterior is obtained as the conjugate distribution of the \textit{Multinomial} likelihood and given by
\begin{equation}\label{diric2}
p(\underline{\pi} \mid \underline{x} ; \eta) = \Gamma (n+k\eta) \prod_{j=1}^{k} \dfrac{\pi_{j}^{x_j+\eta-1}}{\Gamma(x_j+\eta)}.
\end{equation}
\bigskip\\
\textit{Marginal Likelihood.} In order to specify the value of the hyperparameter $\eta$, we estimate this quantity from the maximization of the marginal likelihood which is obtained through the compound \textit{Dirichlet-Multinomial} integration \citep{mosima:1962} 
\begin{eqnarray}\label{mlik}
L(\eta;\underline{x},n) &=& \int_{\Delta_{k-1}} p(\underline{x} \mid \underline{\pi},n) p(\underline{\pi}; \eta) d_{k-1}\underline{\pi} \nonumber\\
 &=& \dfrac{\Gamma(n+1)\Gamma(k\eta)}{\Gamma(n+k\eta)} \prod_{j=1}^{k}\dfrac{\Gamma(x_j+\eta)}{\Gamma(x_j+1)\Gamma(\eta)}I(\Sigma_j x_j-n).
\end{eqnarray}
Notice that the integration in the above equation is defined over the simplex $\Delta_{k-1}$ with the respective integrating measure that we denoted by $d_{k-1}\underline{\pi}$ for simplicity of notation. Then, a simple solution $\hat{\eta}$, approximating the maximization of Equation \eqref{mlik}, is obtained by the Newton-Raphson iterative equation 
$$
\eta^{(t+1)}=\eta^{(t)}-\dfrac{l'(\eta;\underline{x},n)}{l''(\eta;\underline{x},n)},
$$
where
$$
l'(\eta;\underline{x},n)=-\sum_{m=0}^{n-1} \dfrac{k}{k\eta+m} + \sum_{j=1}^k\sum_{y=0}^{x_j-1}\dfrac{1}{\eta+y},
$$
and
$$
l''(\eta;\underline{x},n)=\sum_{m=0}^{n-1} \dfrac{k^2}{(k\eta+m)^2} - \sum_{j=1}^k\sum_{y=0}^{x_j-1}\dfrac{1}{(\eta+y)^2},
$$
are the first and second derivatives of the log-likelihood $l(\eta;\underline{x},n)=\log L(\eta;\underline{x},n) $ respectively. Notice that the second summation in the second term of both first and second derivatives is cancelled and skipped in the case of $x_j=0$.\bigskip\\
\textit{Empirical Bayes estimation.} Then, by plugging the value $\hat{\eta}$ into Equation \eqref{diric1}, the empirical Bayes estimate of each proportion $\pi_j$ is obtained from Equation \eqref{diric2} by the respective posterior mean
\begin{equation}\label{mean}
\hat{\pi}_j= \dfrac{x_j+\hat{\eta}}{n+k\hat{\eta}},
\end{equation}
for every $j=1,...,k$.
\bigskip\\
\textit{Methodological contribution.} The concentration parameter $\eta$ represents the level of uncertainty included in the \textit{Dirichlet} prior necessary to account for potential overdispersion in the data. The contribution of the present work concerns the ability to estimate $\eta$ from only one sample, although $\eta$ plays the role of dispersion parameter as it controls the marginal variance of each $\pi_j$ derived from Equation \eqref{diric1} 
$$
Var[\pi_j]=\dfrac{k-1}{k^2(1+k\eta)}.
$$
By adopting a symmetric \textit{Dirichlet} prior, the taxonomic labels $c_1,...,c_k$ are \textit{a priori} exchangeable and the marginal likelihood in Equation \eqref{mlik} results composed by the exchangeable components $\dfrac{\Gamma(x_j+\eta)}{\Gamma(x_j+1)\Gamma(\eta)}$ with respect to the index $j$. Therefore, marginally, the values $x_1,...,x_k$ are similar to the realizations of exchangeable and identically distributed variables, and the estimate $\hat{\eta}$ represents the average level of concentration obtained from the sampling information with the compound \textit{Dirichlet-Multinomial} likelihood, assuming the noninformative symmetric \textit{Dirichlet} prior to be valid.
\section{Simulation experiments}\label{simul}
\begin{figure}[H]
	\centering
	\includegraphics[scale=0.5]{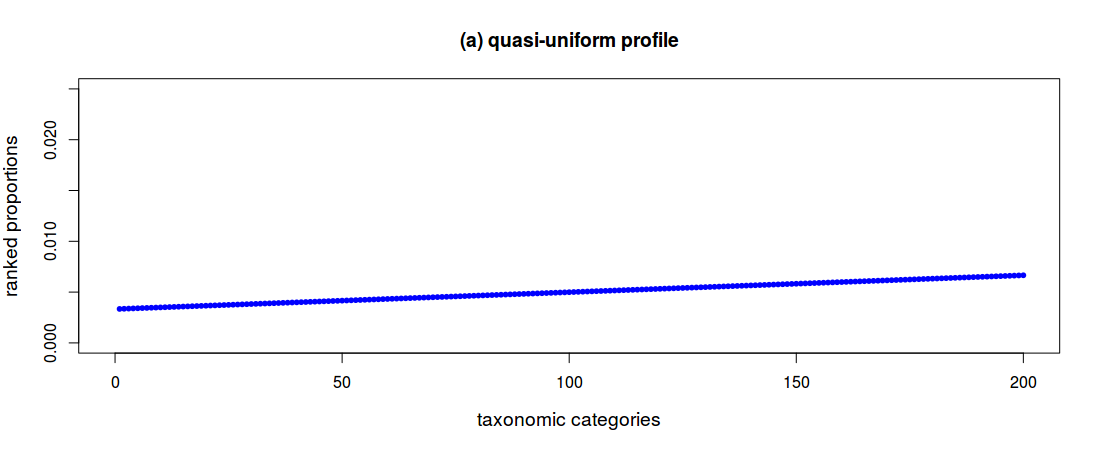}
	\includegraphics[scale=0.5]{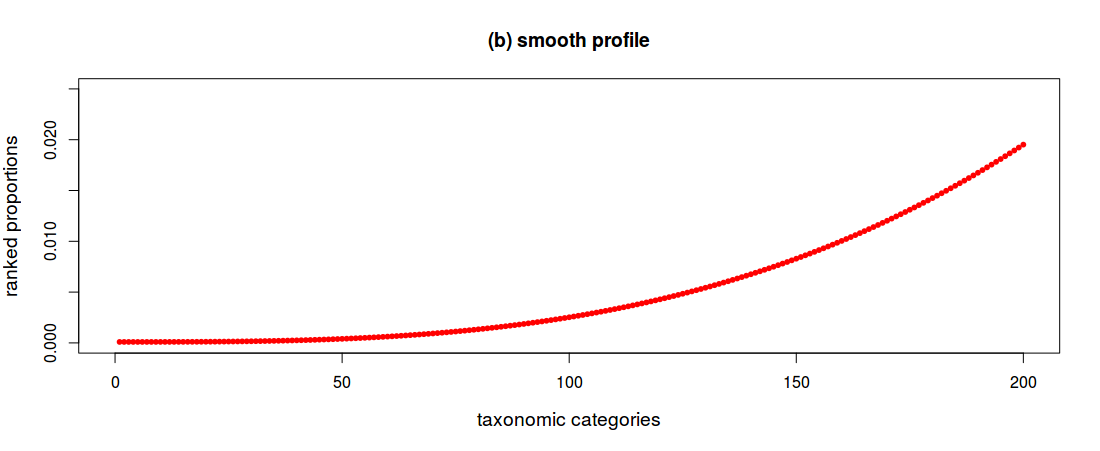}
	\includegraphics[scale=0.5]{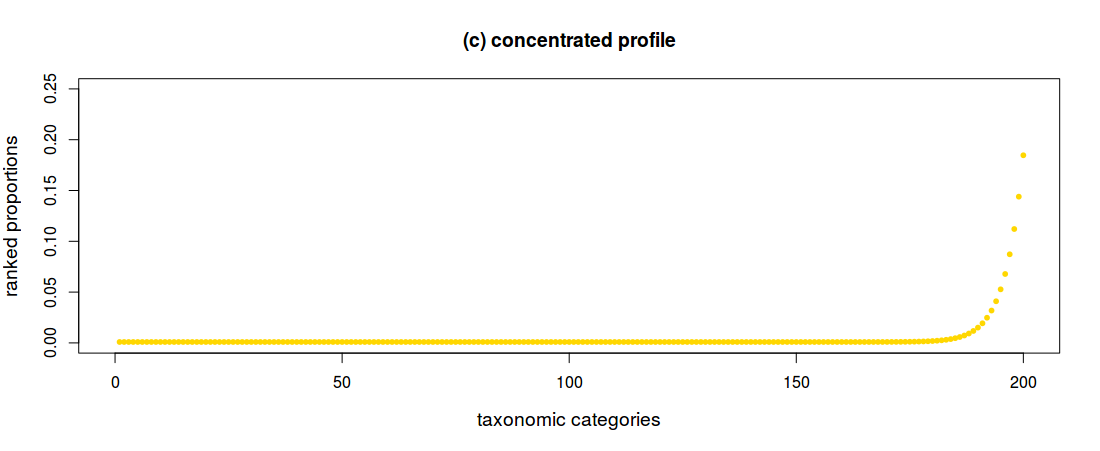}
	\caption{Community theoretical profiles considered in terms of ranked proportions (y-axis) across the categories $c_j$, $j=1,...,k$ (x-axis): (a) quasi-uniform evenness, (b) smooth evenness, and (c) concentrated evenness.}\label{fig:profile}
\end{figure}

In order to investigate the performance of our approach, we present an extensive simulation study in which we compare the empirical Bayes (EB) estimation presented in Section \ref{ebayes} to the classical maximum likelihood (ML) method based on the \textit{Multinomial} model. In particular, we compare the sampling distributions of diversity and similarity indices when the two aforementioned methods are used.\\ 
We build a simulation model to produce samples with varying amount of overdispersion, estimate the proportions of each category from the samples using the EB and ML methods, and calculate diversity and similarity indices based on these estimates. Our goal is to discover under what circumstances the estimators yield the best results, i.e. sampling index values are closest to the population index value calculated with respect to the community true composition.\\
Our simulation model is based on the setup of Section \ref{over}: in each simulation, for every category $c_j$, the respective count $x_j$ is sampled from the \textit{Poisson} distribution in Equation \eqref{pois1} with parameter $\lambda \pi_j$. To include overdispersion in data, the parameters $\lambda$ and $\underline{\pi}$ are considered as random variables and generated from appropriate probability distributions. In particular, the expected sample size $\lambda$ is simulated from the \textit{Gamma} distribution 
\begin{equation}\label{gamma}
p(\lambda;\alpha,\beta)=\dfrac{{\beta}^{\alpha}}{\Gamma(\alpha)} \lambda^{\alpha-1} e^{-\beta \lambda},
\end{equation}
with the parameters $\alpha$ and $\beta$ controlling the level of sample size and the amount of overdispersion due to clustering. Respectively, the composition vector $\underline{\pi}$ is simulated from the \textit{Dirichlet} distribution with parameter $\underline{\theta}=k \gamma \underline{\pi}^*$ 
\begin{equation}\label{diric3}
p(\underline{\pi} ; \underline{\theta}) = \Gamma (\Sigma_j\theta_j) \prod_{j=1}^{k} \dfrac{\pi_{j}^{\theta_j-1}}{\Gamma(\theta_j)},
\end{equation}
where $\underline{\pi}^*=(\pi^*_1,...,\pi^*_k)$ is the true composition and $\gamma$ is the concentration parameter controlling the amount of overdispersion due to heterogeneity. Notice that the concentration parameter is scaled by $k$ simply to better handle simulated data in situations with different number of categories ($k$) but with the same level of overdispersion ($\gamma$).\\
We consider three types of community profile $\underline{\pi}^*$, representing different patterns of taxonomic composition commonly observed in real applications: \textit{quasi-uniform evenness}, \textit{smooth evenness} and \textit{concentrated evenness}. Without loss of generality, we define the three patterns in terms of the quantity $\pi^*_j$ increasingly ranked with respect to $j$, with $j=1,...,k$. Therefore, $c_j$ represents the $j$-th less populated taxonomic category with the true proportion $\pi_j^*$. In detail, the quasi-uniform evenness describes a community where all taxonomic categories are approximately equally common, with the respective ranked value $\pi_j^*$ defined in terms of a linear function (almost flat) of the quantity $j/k$ normalized to unity. Second, the smooth evenness represents a set of proportions gradually increasing, with the ranked $\pi_j^*$ as a cubic function of $j/k$ normalized to unity. Third, the concentrated evenness describes communities where there are a few dominant categories and the rest are rare, with the ranked $\pi_j^*$ as a linear function of $(j/k)^{50}$ normalized to unity. The three evenness patterns are shown in Figure \ref{fig:profile} respectively.\\
\begin{center}
	<Figure \ref{fig:profile}>
\end{center}
In our simulation experiments, we investigate varying amounts of overdispersion due heterogeneity (controlled by $\gamma$) and due to clustering (controlled by $\beta$, given $\alpha$) with respect to different levels of sample size (controlled by $\alpha$). In summary, for each class of experiments, our simulation model can be represented by the following general scheme:
\begin{itemize}
	\item \texttt{set} the number of categories $k$ and \texttt{set} the number of samples $m$
	\item \texttt{set} the true vector $\underline{\pi}^*$ in accordance to $k$;
	\item \texttt{set} the parameters $\alpha$, $\beta$, and $\gamma$ and \texttt{set} $\underline{\theta}=k \gamma \underline{\pi}^*$;
	\item \texttt{for each sample} (from $1$ to $m$):
	\begin{itemize}
		\item \texttt{simulate} $\lambda$ from the \textit{Gamma} in Equation \eqref{gamma};
		\item \texttt{simulate} $\underline{\pi}$ from the \textit{Dirichlet} in Equation \eqref{diric3};
		\item \texttt{simulate} $x_j$ from the \textit{Poisson} in Equation \eqref{pois1}, for every $j=1,...,k$;
	\end{itemize}
	\item \texttt{stop}.
\end{itemize}
Each class of experiments corresponds to a specific scenario and setting of the parameters $\alpha$, $\beta$, and $\gamma$. In detail, we consider three levels of overdispersion due to heterogeneity: high, medium, and low with the corresponding $\gamma$ equal to 1, 10, and 100 respectively; and only one intermediate level of overdispersion due to clustering, with $\beta=0.1$. Combined with the three values 20, 50, and 100 for $\alpha$; this gives the expected sample size $\lambda$ equal to 200, 500, and 1000 respectively. For each class of experiments, we consider $m=1000$ samples, while the number of taxonomic categories is fixed to $k=200$ in order to represent situations with a high but biologically often common level of richness.\\
After data are simulated, for each class of experiment and for each sample in that class, we compute the estimates of the true composition profile $\underline{\pi}^*$ with the EB method and the ML method, obtaining the two vectors $\underline{\hat{\pi}}_{EB}$ (using Equation \eqref{mean}) and $\underline{\hat{\pi}}_{ML}$ (using Equation \eqref{mle}) respectively. Afterwards, we use these estimates to assess diversity and similarity in the simulated samples. Notice that the simulation procedure and the Bayesian model used in the empirical Bayes estimation are different as the simulation model includes two sources of overdispersion, clustering and heterogeneity.\\
For diversity indices, we consider the Shannon entropy \citep{shawea:1963} and the Simpson diversity \citep{simpso:1949}. The Shannon entropy 
$$H=-\sum_{j=1}^k \hat{\pi}_j\log{\hat{\pi}_j}, \; \text{with } H\in [ 0;\log k ],$$
originating from information theory \citep{shawea:1963} remains one of the most popular diversity indices used in ecology \citep{magurr:2004} while the Simpson diversity index 
$$D=\sum_{j=1}^k \hat{\pi}_j^2, \; \text{with } D\in [ 0;1/k],$$ 
is one of the most robust diversity measures and gives more weight to the most abundant species in the sample \citep{magurr:2004}. To study how overdispersion affects the assessment of similarity, we use the percentage similarity of \citet{renkon:1938}, which \citet{novbod:1992} refer to as the percent model affinity (PMA). The PMA is typically used to measure the similarity of two samples but here we use the respective PMA index to measure the similarity of the estimated profile $\underline{\hat{\pi}}$ ($\underline{\hat{\pi}}_{EB}$ or $\underline{\hat{\pi}}_{ML}$) with respect to the true profile $\underline{\pi}^*$, that is 
$$I=1-\frac{1}{2}\sum_{j=1}^k |\hat{\pi}_j-\pi^*_j|, \; \text{with } I\in [ 0;1].$$ 
While the PMA index $I$ uses the absolute difference between two profiles, the Euclidean similarity is calculated with the squared Euclidean distance \citep{cliste:1975} 
$$E=1-\sum_{j=1}^k (\hat{\pi}_j-\pi^*_j)^2, \; \text{with } E\in [-1;1].$$
For every index ($H$, $D$, $I$, and $E$), one estimation method outperforms the other when the respective sampling distribution of index values tends to be closer to the true index value, i.e. when the respective sampling variation combined with the level of bias is smaller. This information is presented in Tables 1--4 where the respective sampling mean, standard deviation (SD), bias, and root mean squared error (RMSE) of the empirical Bayes and maximum likelihood estimates are reported with respect to the different simulation scenarios and composition profiles. Using the numerical results and the graphical boxplots shown in Figures \ref{fig:boxplot1}--\ref{fig:boxplot4} (the respective numerical quantiles are reported in Appendix A), we compare the performance of the empirical Bayes estimation to the maximum likelihood estimation. In those Tables 1--4, for each composition profile $\underline{\pi}^*$ (quasi-uniform, smooth, concentrated), we denote by $H^*$, $D^*$, $I^*$, and $E^*$ the respective true value of each index ($H$, $D$, $I$, and $E$).\\
\begin{center}
	<Figure \ref{fig:boxplot1}>
\end{center}
\begin{center}
	<Figure \ref{fig:boxplot2}>
\end{center}
\begin{center}
	<Figure \ref{fig:boxplot3}>
\end{center}
\begin{center}
	<Figure \ref{fig:boxplot4}>
\end{center}
\begin{center}
	<Table \ref{mse1}>
\end{center}
\begin{center}
	<Table \ref{mse2}>
\end{center}
\begin{center}
	<Table \ref{mse3}>
\end{center}
\begin{center}
	<Table \ref{mse4}>
\end{center}
\textit{Shannon entropy} (Figure \ref{fig:boxplot1} and Table \ref{mse1}). In general, the EB method performs better than the ML method and results in a less biased estimation with a smaller sampling RMSE. As expected, when the expected sample size increases ($\alpha$ grows) or the level of overdispersion due to heterogeneity declines ($\gamma$ increases) the ML approaches the EB estimation. ML shows a general tendency to underestimate the Shannon entropy while the EB overestimates it, but EB has a better level of unbiasedness. The variation of both methods is larger for the concentrated composition, however it decreases with increased expected sample size for all the compositions. For the quasi-uniform evenness, the EB always outperforms the ML estimation, and significantly so when the overdispersion is high ($\gamma=1$) in samples of small size ($\alpha=20$). For the smooth profile, the EB method tends to overestimate the entropy when the level of overdispersion decreases ($\gamma=10,100$) in samples of small and intermediate size ($\alpha=20,50$), but retains a better level of unbiasedness when compared to the ML method. Further, with the smooth composition, the EB method performs significantly better when the overdispersion is high ($\gamma=1$), independently of the sample size. For the concentrated profile, the superior performance of the EB estimation is also evident as it is always less biased than ML. \citet{tongyl:1983} studied the properties of the Shannon entropy under the \textit{Multinomial} model and showed that its bias vanishes when the sample size increases (Theorem 3). We observed this phenomenon when there was low overdispersion in the data ($\gamma=100$), with the simulated model generating samples more similar  to \textit{Multinomial} counts. \bigskip\\
\textit{Simpson diversity} (Figure \ref{fig:boxplot2} and Table \ref{mse2}). In general, the behaviour of the Simpson index is quite similar to the behaviour of the Shannon entropy. The main difference is that the ML method tends to overestimate the true value $D^*$. In the cases of quasi-uniform and smooth profiles, the same observations made about the Shannon entropy are also valid for the Simpson index, with the EB estimation generally outperforming the ML estimation. For the concentrated evenness, the EB method outperforms the ML method only when the level of overdispersion is high ($\gamma=1$), independently of the sample size. When the overdispersion is intermediate or low ($\gamma=10,100$), the EB method underestimates the true value $D^*$ while the ML method produces less biased results and outperforms EB, especially  with samples of intermediate and large size ($\alpha=50,100$). For these 6 scenarios the maximum likelihood sampling RMSE is smaller than the empirical Bayes sampling RMSE. Further, when the profile is concentrated, both procedures produce estimates with larger variation. \bigskip\\
\textit{PMA index} (Figure \ref{fig:boxplot3} and Table \ref{mse3}). Also for the PMA index, EB performs generally better than ML, although both methods heavily underestimate the true value $I^*=1$ (perfect similarity) in some cases. This behaviour is expected as it is due to the strong bias affecting the estimation of the PMA index in cases of near perfect similarity \citep{arjeal:2016}. For the quasi-uniform evenness, the superiority of the EB method is particularly obvious with a significant difference between the two methods. Also for both the smooth and concentrated profiles, the EB performs generally better than ML providing more unbiased estimates in almost all situations. But in these cases the ML approaches the EB performance, with levels of the respective sampling RMSE which are comparable. Empirical results for the simulation model closer to the \textit{Multinomial} type model ($\gamma=100$) are supported by the asymptotics presented in \cite{arjeal:2016} (Theorem 1). \bigskip\\
\textit{Euclidean similarity} (Table \ref{mse4} and Figure \ref{fig:boxplot4}). For the Euclidean similarity, the difference between the two methods is marginal and the estimators are both unbiased although numerical values presented in Table \ref{mse4} suggest a slightly improved performance of the EB method, especially for the quasi-uniform profile. The values of the sampling RMSE are always significantly small for both methods. The stability of the Euclidean similarity with regards to the model used was observed also in a study by \citet{arjeal:2017} that used simulations to assess the effect of classification errors on the values of different indices.\\
\begin{center}
	<Table \ref{eff}>
\end{center}
In Table \ref{eff}, the relative efficiency of the EB estimation compared to the ML estimation is reported with respect to the different simulation scenarios and profiles. As an indicator of efficiency, we consider the sampling ratio between the inverse of the empirical Bayes root mean squared error (denoted by $RMSE_{EB}$) and the inverse of the maximum likelihood root mean squared error (denoted by $RMSE_{ML}$)
$$
\text{Eff}_{EB/ML} = \dfrac{RMSE_{ML}}{RMSE_{EB}}.
$$
In the last column of Table \ref{eff}, we report also the relative efficiency computed (partially) for each profile and (totally) for each indicator (see the Appendix B for details). The relative efficiency $\text{Eff}_{EB/ML}$ allows us to quantify the superiority($>1$) or inferiority ($<1$) of the EB method with respect to the ML method. When $\text{Eff}_{EB/ML}$ is close to 1, the two methods are equivalent in terms of efficiency.\\  
For the Shannon entropy, EB gets the highest level of superiority with the quasi-uniform evenness when samples are small ($\alpha=20$) and the level of overdispersion is intermediate ($\gamma=10$). In this case, the relative efficiency is equal to $42.4$, that is the EB estimation is around 42 times more efficient than the ML estimation. For the Shannon entropy, the EB is more efficient than ML for all the three profiles, with values of $\text{Eff}_{EB/ML}$ equal to 3.2, 2.1, and 1.9 respectively.\\ 
For the Simpson diversity, independently of the sample size, the ML is superior to the EB for the concentrated profile when overdispersion is not high. In these 6 situations, the EB performs with 60\% to 80\% of the ML efficiency and for this profile, EB is 11\% less efficient than ML. In all the other situations EB is superior with the relative efficiency $\text{Eff}_{EB/ML}$ between 1.1 and 37.3. For the quasi-uniform and smooth profiles, EB is more efficient with values of $\text{Eff}_{EB/ML}$ equal to 2.7 and 2.1 respectively.\\
For the PMA index, in some situations the two methods perform at the same level of efficiency (when $\text{Eff}_{EB/ML}$ is close to 1) but the EB is more efficient than ML for all the profiles, with an increase in efficiency of 90\% ($\text{Eff}_{EB/ML}=1.9$), 10\% ($\text{Eff}_{EB/ML}=1.1$), and 10\% ($\text{Eff}_{EB/ML}=1.1$) respectively.\\
For the Euclidean similarity, the superiority of the EB estimation is evident when the profile is quasi-uniform or smooth with $\text{Eff}_{EB/ML}$ equal to 2.6 and 1.5 respectively, while for the concentrated profile the two methods present approximately the same level of efficiency with $\text{Eff}_{EB/ML}$ equal to 1.1.\\
In summary, the EB method performs generally better than the ML method in both the assessments of diversity and similarity. The relative efficiency $\text{Eff}_{EB/ML}$ for each indicator is equal to 2.4 (Shannon entropy $H$), 1.1 (Simpson diversity $D$), 1.4 (PMA index $I$), and 1.4 (Euclidean similarity $E$); with an increase in efficiency of 140\%, 10\%, 40\%, and 40\% respectively.
\section{An application to real data} \label{appl}
In the simulation study above, the true underlying species profile that produces the samples is known. With this knowledge, we can objectively assess the best estimation method and evaluate how much variation and bias is created in the index values due to different estimators of species proportions. However in practice, the true profile is not known. When sampling a waterbody for biomonitoring and targeting, e.g. macroinvertebrates or diatoms, the use of all the information contained within one sample can be compared through the index values obtained with the different methods. \\
To study the difference between the proposed empirical Bayes estimation and the classical \textit{Multinomial} maximum likelihood estimation by using real data, we utilize an available set of Finnish data with information on the taxonomic composition of benthic freshwater macroinvertebrates. The data have been collected in the national biomonitoring program for purposes of ecological status assessment under the European Union Water Framework Directive (\texttt{http://data.europa.eu/eli/dir/2000/60/oj} for details) during 2006–2012. We use a set of 484 samples belonging to 12 Finnish stream types categorized according to their location (southern, northern), their size (small, medium, or large) and soil typology (peatland or woodland). Further, each stream type has samples from both reference sites considered to be unimpacted by human actions and impacted sites, resulting in a total of 24 stream categories. The number of samples for each individual stream category varies from 3 to 129. Sample sizes display a great deal of variation. For the reference sites, the sample size varies from 19 to 13820 individuals with an average sample size of 1367. For the impacted sites, the sample size varies from 40 to 11780 individuals with an average sample size of 1444. With this data, we expect that the differences between methods are small due to large sample sizes, fortunately decreasing the effect of overdispersion. \\
Before the use of the empirical Bayes estimator, we need to know or fix the number of taxonomic classes. We listed all the species found within each of the 24 stream types and assumed that if a species was found once within one stream type, it could be present in any of the sampled communities of that type. As a result, the number of categories for the reference streams varies from 46 to 103 and for the impacted streams from 45 to 123. After having the value for $k,$ we estimate the species proportions for each sample with the empirical Bayes method and the maximum likelihood method, and calculate the index values to study if there are differences between the two estimators. Since the true reference profile needed for the calculation of the PMA index $I$ and the Euclidean similarity $E$ is not known with real data, the comparison is restricted to the Shannon entropy $H$ and the Simpson index $D,$ separately for the reference and impacted samples. \\  
In Figures 7--10, the effects of the sample size and the overdispersion on the estimates of indices are compared. When considering the Shannon entropy for reference and impacted sites, we find that the EB results in slightly higher estimates than the ML method (Figures 7--8). The range of average differences between estimators are (0.024, 0.198) with mean equal to 0.072 and standard deviation equal to 0.046  for reference sites, and (0.029, 0.144) with mean equal to 0.072 and standard deviation equal to 0.032 for impacted sites. As expected, the estimates of indices are more similar when the sample size is larger (Figures 7--8). For the Simpson index, the estimates obtained with EB and ML are almost equal, except for some samples with very small size (Figures 9--10). The range of average differences between estimators are (-0.010, -0.001) with mean equal to -0.004 and standard deviation equal to 0.002 for reference sites, and (-0.007, -0.002) with mean equal to -0.004 and standard deviation equal to 0.002 for impacted sites. The range of $\hat\eta$ for the reference sites is 3.3-17.4 and 1.6-18.2 for impacted sites. In the simulation experiment with an intermediate level of overdispersion due to heterogeneity ($\gamma$ = 10), the EB method was always less biased than ML for the Shannon entropy. For the Simpson diversity, the EB and ML methods produced estimates quite equal with the same level of overdispersion, when the profile is smooth and sample size large, and when the profile is concentrated.
\begin{center}
	<Figure \ref{fig:app1}>
\end{center}
\begin{center}
	<Figure \ref{fig:app2}>
\end{center}
\begin{center}
	<Figure \ref{fig:app3}>
\end{center}
\begin{center}
	<Figure \ref{fig:app4}>
\end{center}
\section{Conclusion} \label{final}
The assessment of community diversity and the evaluation of similarity between communities is of great interest when monitoring the status of ecosystems. All current diversity and similarity indicators are based on the estimates of the taxonomic composition of such communities computed from samples and tacitly assume the classical \textit{Poisson-Multinomial} model. But there is a serious and realistic question: How does the potential overdispersion present in the available data affect our results?\\ 
Overdispersion in the response variable such as taxonomic counts may be the result of the action of significant environmental factors, with effects of clustering within the community and heterogeneity in its composition. In the current setting, we assumed that these factors are unobservable, which is often true with real data applications. Then, following the empirical Bayes approach combined with the marginal likelihood method, we presented a Bayesian model accounting the presence of overdispersion with a novel estimation method of the taxonomic proportions. We used a noninformative symmetric \textit{Dirichlet} prior combined with the \textit{Multinomial} likelihood. The concentration parameter of the \textit{Dirichlet} prior was estimated using the respective compound \textit{Dirichlet-Multinomial} model while the taxonomic proportions were estimated as the Bayesian posterior means.\\ 
The methodologically relevant novelty of our proposal is the ability to jointly estimate both the taxonomic proportions and the concentration parameter with only one sample. The important assumption underlying our method is the validity of the symmetric prior that allows us to consider the data similar to realizations of exchangeable variables, with the ability to estimate an average level of overdispersion between the taxonomic counts collected from one sample. An advantage of the adopted empirical Bayes approach is that it is a data driven method with a computationally easy and fast solution based on the Newton-Raphson optimization. The only problem in our applications is the potential fault of the Newton-Raphson algorithm when the marginal likelihood is significantly flat, i.e. the observed counts are too similar to each other and do not include the necessary statistical information to get a numerical approximation of the parameter $\eta$. This situation may occur with data from a community with complete uniform evenness and low overdispersion but is extremely unlikely to happen with samples of real biological communities.\\
Our proposal was evaluated through a simulation study with four indicators typically used in biodiversity assessment of ecosystems: the Shannon entropy, the Simpson diversity, the percent model affinity, and the Euclidean similarity. The performance of our approach was compared to the classical maximum likelihood method in terms of sampling distribution of each estimated index when matched to the respective true value. In the simulation experiments, we considered three types of community composition (quasi-uniform, smooth, and concentrated) related to 200 taxonomic categories, three levels of overdispersion due to heterogeneity, one level of overdispersion due to clustering, and three values of expected sample size.\\ 
From the results, we can assert that the proposed empirical Bayes estimation generally performs better than the classical maximum likelihood method in both the assessments of diversity and similarity. In some situations the ML estimation approaches the EB estimation (large samples with low overdispersion) and in 6 of 108 cases (in total, we simulated 27 scenarios for each of four indicators) the ML method is more efficient than the EB estimation with the Simpson diversity. The EB method performs significantly better with small samples. This is due to the property of Bayesian procedures to produce parameter estimates which are smooth towards the prior mean. Therefore, when the sampling information is poor, i.e. sample size is small, the empirical Bayes estimation uses the additional information included in the prior distribution model. In fact, the superiority of the empirical Bayes is particularly obvious when the evenness is quasi-uniform. This is expected as the quasi-uniform profile is the closest to the prior choice of symmetric Dirichlet distribution. For the concentrated profile the performance of the two methods are more comparable. However, for the smooth profile, which is the most common situation with real applications, the EB always provides the most efficient estimates.\\
The observed difference in the estimates of biodiversity indices due to overdispersion are generally relatively small, but still point out the importance of model choice. Our simulation study results demonstrate an overall superiority of the EB approach. The better performance of the ML method for the Simpson diversity in some scenarios does not significantly impede on the overall utility of the EB estimation. Arguably in those few cases EB would produce only slightly more conservative estimates unlikely to change the overall  assessment. Thus in conclusion, when using the EB approach as an additional tool, error propagation due to inherent properties of biological data becomes less likely which improves managerial decisions. Therefore, we would recommend practitioners in biomonitoring to use EB which is more efficient with high level of overdispersion in small samples and also handles other situations well. However ML is recommended in situations of low overdispersion in large samples.\\
In both the simulation study and the real data application, we focused on communities characterized by a large number of taxonomic categories, i.e. with a high level of richness, such as diatoms, phytoplankton, aquatic macroinvertebrates, or bacteria. The effects of varying richness on the statistical distribution of diversity and similarity indicators is deferred to subsequent work.\\  
The proposed model is designed for one sample with a fixed sample size such that overdispersion due to heterogeneity can be properly assessed. In our current setting, specific overdispersion due to clustering can be only partially accounted for. Proper consideration of overdispersion of this type would require the extension of our empirical Bayes approach to several samples through the inclusion of a \textit{Gamma-Poisson} component to model the randomness of the parameter $\lambda$. Such a generalization may be one of the challenges in the future. As a further possible extension, the number of taxonomic categories $k$ might be also considered as a random quantity allowing to include into the model an additional source of overdispersion. Furthermore, when the sources of overdispersion are known, the Bayesian scheme can include also the effect of covariates in a \textit{Multinomial} logistic regression framework.
\paragraph{Computation.} The algorithm for the empirical Bayes estimation and the simulation procedures were implemented in \texttt{R}. Codes and simulated data are available from the authors.
\bigskip\\
\paragraph{Acknowledgement.} This joint work is supported by the Academy of Finland (projects no. 289076 (FD, J\"{A}, SK) and no. 289104 (FD, KM)). Fabio Divino is also partially supported by the PRIN2015 project “Environmental processes and human activities: capturing their interactions via statistical methods (EPHASTAT)” funded by MIUR (Italian Ministry of Education, University and Scientific Research) and acknowledges the support of the Department of Mathematics and Statistics of the University of Jyv\"{a}skyl\"{a} and the Finnish Environment Institute SYKE, Jyv\"{a}skyl\"{a}. 

\begin{thebibliography}{}
	
	\bibitem[\"Arje {\em et~al.}(2016)\"Arje, Choi, Divino, Meissner, and
	K\"arkk\"ainen]{arjeal:2016}
	\"Arje, J., Choi, K.-P., Divino, F., Meissner, K., and K\"arkk\"ainen, S.
	(2016).
	\newblock Understanding the statistical properties of the percent model
	affinity index can improve biomonitoring related decision making.
	\newblock {\em Stochastic Environmental Research and Risk Assessment\/}, {\bf
		30}(7), 1981--2008.
	
	\bibitem[\"Arje {\em et~al.}(2017)\"Arje, K\"arkk\"ainen, Meissner, Iosifidis,
	Ince, Gabbouj, and Kiraynaz]{arjeal:2017}
	\"Arje, J., K\"arkk\"ainen, S., Meissner, K., Iosifidis, A., Ince, T., Gabbouj,
	M., and Kiraynaz, S. (2017).
	\newblock The effect of automated taxa identification errors on biological
	indices.
	\newblock {\em Expert Systems with Applications\/}, {\bf 72}, 108--120.
	
	\bibitem[Bates and Neyman(1952)Bates and Neyman]{batney:1952}
	Bates, G.~E. and Neyman, J. (1952).
	\newblock Contributions to the theory of accident proneness. i. an optimistic
	model of the correlation between light and severe accidents.
	\newblock {\em University of California Publications in Statistics\/}, {\bf
		1}(9), 215--254.
	
	\bibitem[Birk {\em et~al.}(2012)Birk, Bonne, Borja, Brucet, Courrat, Poikane,
	Solimini, van~de Bund, Zampoukas, and Hering]{birkal:2012}
	Birk, S., Bonne, W., Borja, A., Brucet, S., Courrat, A., Poikane, S., Solimini,
	A., van~de Bund, W., Zampoukas, N., and Hering, D. (2012).
	\newblock Three hundred ways to assess {E}urope's surface waters: an almost
	complete overview of biological methods to implement the {W}ater {F}ramework
	{D}irective.
	\newblock {\em Ecological Indicators\/}, {\bf 18}, 31--41.
	
	\bibitem[Bliss and Fisher(1953)Bliss and Fisher]{blifis:1953}
	Bliss, C.~I. and Fisher, R.~A. (1953).
	\newblock Fitting the {N}egative {B}inomial distribution to biological data.
	\newblock {\em Biometrics\/}, {\bf 9}(2), 176--200.
	
	\bibitem[Breslow(1984)Breslow]{breslow:1984}
	Breslow, N.~E. (1984).
	\newblock Extra-{P}oisson variation in log-linear models.
	\newblock {\em Journal of the Royal Statistical Society. Series C (Applied
		Statistics)\/}, {\bf 33}(1), 38--44.
	
	\bibitem[Carlin and Louis(2000)Carlin and Louis]{carlou:2000}
	Carlin, B.~P. and Louis, T.~A. (2000).
	\newblock {\em Bayes and {E}mpirical {B}ayes {M}ethods for {D}ata
		{A}nalysis\/}.
	\newblock Chapman \& Hall, London, UK.
	
	\bibitem[Chen and Li(2013)Chen and Li]{chenli:2013}
	Chen, J. and Li, H. (2013).
	\newblock Variable selection for sparse {D}irichlet-{M}ultinomial regression
	with an application to microbiome data analysis.
	\newblock {\em The Annals of Applied Statistics\/}, {\bf 7}(1), 418--442.
	
	\bibitem[Clapham(1936)Clapham]{clapha:1936}
	Clapham, A.~R. (1936).
	\newblock Over-dispersion in grassland communities and the use of statistical
	methods in plant ecology.
	\newblock {\em Journal of Ecology\/}, {\bf 24}(1), 232--251.
	
	\bibitem[Clifford and Stephenson(1975)Clifford and Stephenson]{cliste:1975}
	Clifford, H.~T. and Stephenson, W. (1975).
	\newblock {\em An introduction to numerical classification\/}.
	\newblock London: Academic Press, London, UK.
	
	\bibitem[Cox(1983)Cox]{cox:1983}
	Cox, D.~R. (1983).
	\newblock Some remarks on overdispersion.
	\newblock {\em Biometrika\/}, {\bf 70}(1), 269--274.
	
	\bibitem[de~Valpine and Harmon-Threatt(2013)de~Valpine and
	Harmon-Threatt]{valpha:2013}
	de~Valpine, P. and Harmon-Threatt, A.~N. (2013).
	\newblock General models for resource use or other compositional count data
	using the {D}irichlet-multinomial distribution.
	\newblock {\em Ecology\/}, {\bf 94}(12), 2678--2687.
	
	\bibitem[Diaconis and Ylvisaker(1979)Diaconis and Ylvisaker]{diaylv:1979}
	Diaconis, P. and Ylvisaker, D. (1979).
	\newblock Conjugate priors for exponential family.
	\newblock {\em The Annals of Statistics\/}, {\bf 7}(2), 269--281.
	
	\bibitem[Dodge(2008)Dodge]{dodge:2008}
	Dodge, Y. (2008).
	\newblock {\em The {C}oncise {E}ncyclopedia of {S}tatistics\/}.
	\newblock Springer, New York, US.
	
	\bibitem[Etterson {\em et~al.}(2009)Etterson, Niemi, and Danz]{etteral:2009}
	Etterson, M.~A., Niemi, G.~J., and Danz, N.~P. (2009).
	\newblock Estimating the effects of detection heterogeneity and overdispersion
	on trends estimated from avian point counts.
	\newblock {\em Ecological Applications\/}, {\bf 19}(8), 2049--2066.
	
	\bibitem[Feller(1943)Feller]{feller:1943}
	Feller, W. (1943).
	\newblock On a general class of contagious distributions.
	\newblock {\em The Annals of Mathematical Statistics\/}, {\bf 14}(4), 389--400.
	
	\bibitem[Gotelli and Chao(2013)Gotelli and Chao]{gotcha:2013}
	Gotelli, N.~J. and Chao, A. (2013).
	\newblock Measuring and estimating species richness, species diversity, and
	biotic similarity from sampling data.
	\newblock In S.~A. Levin, editor, {\em Encyclopedia of {B}iodiversity, Volume
		5\/}, pages 195--211. Academic Press, Waltham, US.
	
	\bibitem[Greenwood and Yule(1920)Greenwood and Yule]{greyul:1920}
	Greenwood, M. and Yule, G.~U. (1920).
	\newblock An inquiry into the nature of frequency distributions representative
	of multiple happenings with particular reference to the occurrence of
	multiple attacks of disease or of repeated accidents.
	\newblock {\em Journal of the Royal Statistical Society\/}, {\bf 83}(2),
	255--279.
	
	\bibitem[Harrison(2014)Harrison]{harris:2014}
	Harrison, X.~A. (2014).
	\newblock Using observation-level random effects to model overdispersion in
	count data in ecology and evolution.
	\newblock {\em PeerJ\/}, (2:e616), https://doi.org/10.7717/peerj.616.
	
	\bibitem[Hilbe(2011)Hilbe]{hilbe:2011}
	Hilbe, J.~M. (2011).
	\newblock {\em Negative {B}inomial {R}egression\/}.
	\newblock Cambridge University Press, Cambridge, UK.
	
	\bibitem[Kokonendji(2014)Kokonendji]{kokone:2014}
	Kokonendji, C.~C. (2014).
	\newblock Over- and underdispersion models.
	\newblock In N.~Balakrishnan, editor, {\em Methods and {A}pplications of
		{S}tatistics in {C}linical {T}rials, {V}olume 2: {P}lanning, {A}nalysis, and
		{I}nferential {M}ethods\/}, pages 506--526. John Wiley \& Sons, New York, US.
	
	\bibitem[Levy and Lemeshow(2008)Levy and Lemeshow]{levlem:2008}
	Levy, P.~S. and Lemeshow, S. (2008).
	\newblock {\em Sampling of {P}opulations\/}.
	\newblock John Wiley \& Sons, New York, US.
	
	\bibitem[Linden and Mäntyniemi(2011)Linden and Mäntyniemi]{linman:2011}
	Linden, A. and Mäntyniemi, S. (2011).
	\newblock Using the negative binomial distribution to model overdispersion in
	ecological count data.
	\newblock {\em Ecology\/}, {\bf 92}(7), 1414--1421.
	
	\bibitem[Lindsey(1995)Lindsey]{lindse:1995}
	Lindsey, J.~K. (1995).
	\newblock {\em Modelling {F}requency and {C}ount {D}ata\/}.
	\newblock Oxford University Press, Oxford, UK.
	
	\bibitem[Magurran(2004)Magurran]{magurr:2004}
	Magurran, A.~E. (2004).
	\newblock {\em Measuring {B}iological {D}iversity\/}.
	\newblock Balckwell, Oxford, UK.
	
	\bibitem[Manly and Navarro~Alberto(2015)Manly and Navarro~Alberto]{mannav:2015}
	Manly, B. F.~J. and Navarro~Alberto, J.~A. (2015).
	\newblock {\em Introduction to {E}cological {S}ampling\/}.
	\newblock CRC Press, London, UK.
	
	\bibitem[Matthews and Whittaker(2015)Matthews and Whittaker]{matwhi:2015}
	Matthews, T.~J. and Whittaker, R.~J. (2015).
	\newblock On the species abundance distribution in applied ecology and
	biodiversity management.
	\newblock {\em Journal of Applied Ecology\/}, {\bf 52}, 443--454.
	
	\bibitem[McCullagh and Nelder(1983)McCullagh and Nelder]{mccnel:1983}
	McCullagh, P. and Nelder, J.~A. (1983).
	\newblock {\em Generalized linear models (1. ed.)\/}.
	\newblock Chapman \& Hall, London.
	
	\bibitem[Mosimann(1962)Mosimann]{mosima:1962}
	Mosimann, J.~E. (1962).
	\newblock On the compound multinomial distribution, the multivariate
	$\beta$-distribution, and correlations among proportions.
	\newblock {\em Biometrika\/}, {\bf 49}(1/2), 65--82.
	
	\bibitem[Nelson(1984)Nelson]{nelson:1984}
	Nelson, J.~F. (1984).
	\newblock The {D}irichlet-{G}amma-{P}oisson model of repeated events.
	\newblock {\em Sociological Methods \& Research\/}, {\bf 12}(4), 347--373.
	
	\bibitem[Nelson(1985)Nelson]{nelson:1985}
	Nelson, J.~F. (1985).
	\newblock Multivariate {G}amma-{P}oisson models.
	\newblock {\em Journal of the American Statistical Association\/}, {\bf
		80}(392), 828--834.
	
	\bibitem[Novak and Bode(1992)Novak and Bode]{novbod:1992}
	Novak, M.~A. and Bode, R.~W. (1992).
	\newblock Percent model affinity: a new measure of macroinvertebrate community
	composition.
	\newblock {\em Journal of the North American Benthological Society\/}, {\bf
		11}(1), 80--85.
	
	\bibitem[Patil(1962)Patil]{patil:1962}
	Patil, G.~P. (1962).
	\newblock On certain compound {P}oisson and compound {B}inomial distributions.
	\newblock {\em Sankhya, A\/}, {\bf 26}(2/3), 293--294.
	
	\bibitem[Patil(1965)Patil]{patil:1965}
	Patil, G.~P. (1965).
	\newblock Certain characteristic properties of multivariate discrete
	probability distributions akin to the {B}ates-{N}eyman model in the theory of
	accident proneness.
	\newblock {\em Sankhya, A\/}, {\bf 27}(2/4), 259--270.
	
	\bibitem[Pfeifer {\em et~al.}(1998)Pfeifer, Bräumer, Dekker, and
	Schleier]{pfebau:1998}
	Pfeifer, D., Bräumer, H.-P., Dekker, R., and Schleier, U. (1998).
	\newblock Statistical tools for monitoring benthic communities.
	\newblock {\em Senckenbergiana maritima\/}, {\bf 12}, 63--76.
	
	\bibitem[Pielou(1969)Pielou]{pielou:1969}
	Pielou, E.~C. (1969).
	\newblock {\em An {I}ntroduction to {M}athematical {E}cology\/}.
	\newblock John Wiley \& Sons, New York, US.
	
	\bibitem[Pielou(1975)Pielou]{pielou:1975}
	Pielou, E.~C. (1975).
	\newblock {\em Ecological {D}iversity\/}.
	\newblock John Wiley \& Sons, New York, US.
	
	\bibitem[Qian and Cuffney(2014)Qian and Cuffney]{qianal:2014}
	Qian, S.~S. and Cuffney, T.~F. (2014).
	\newblock A hierarchical zero-inflated model for species compositional
	data-from individual taxon responses to community response.
	\newblock {\em Limnology and Oceanography: Methods\/}, {\bf 12}(7), 498--506.
	
	\bibitem[Rao(1980)Rao]{rao:1980}
	Rao, C.~R. (1980).
	\newblock Diversity and dissimilarity coefficients: a unified approach.
	\newblock {\em Theoretical Population Biology\/}, {\bf 21}, 24--43.
	
	\bibitem[Renkonen(1938)Renkonen]{renkon:1938}
	Renkonen, O. (1938).
	\newblock {Statistische-ökologische Untersuchungen über die terrestrische
		Käferwelt der finnischen Bruchmoore}.
	\newblock {\em Ann. Zool. Soc. Bot. Fenn. Vanamo\/}, {\bf 6}, 1--231.
	
	\bibitem[Richards(2008)Richards]{richar:2008}
	Richards, S.~A. (2008).
	\newblock Dealing with overdispersion count data in applied ecology.
	\newblock {\em Journal of Applied Ecology\/}, {\bf 45}, 218--227.
	
	\bibitem[Shannon and Weaver(1963)Shannon and Weaver]{shawea:1963}
	Shannon, C. and Weaver, W. (1963).
	\newblock {\em The mathematical theory of communication\/}.
	\newblock University Illinois Press, Urbana, USA.
	
	\bibitem[Simpson(1949)Simpson]{simpso:1949}
	Simpson, E.~H. (1949).
	\newblock Measurement of diversity.
	\newblock {\em Nature\/}, {\bf 163}, 688.
	
	\bibitem[Spellerberg(2005)Spellerberg]{spelle:2005}
	Spellerberg, I. (2005).
	\newblock {\em Monitoring {E}cological {C}hange\/}.
	\newblock Cambridge University Press, Cambridge, UK.
	
	\bibitem[Stehman and Overton(1994)Stehman and Overton]{steove:1994}
	Stehman, S.~V. and Overton, W.~S. (1994).
	\newblock Environmental sampling and monitoring.
	\newblock In P.~Patil and C.~R. Rao, editors, {\em Handbook of {S}tatistics,
		Volume 12\/}, pages 263--306. Elsevir, New York.
	
	\bibitem[Thompson(1992)Thompson]{thomps:1992}
	Thompson, S.~K. (1992).
	\newblock {\em Sampling\/}.
	\newblock John Wiley \& Sons, London, UK.
	
	\bibitem[Tong(1983)Tong]{tongyl:1983}
	Tong, Y.~L. (1983).
	\newblock Some distribution properties of the sample species-divervsity indices
	and their applications.
	\newblock {\em Biometrics\/}, {\bf 39}(4), 999--1008.
	
	\bibitem[Wang {\em et~al.}(2012)Wang, Stein, Gao, and Ge]{wangal:2012}
	Wang, J.-F., Stein, A., Gao, B.-B., and Ge, Y. (2012).
	\newblock A review of spatial sampling.
	\newblock {\em Spatial Statistics\/}, {\bf 2}, 1--14.
	
	\bibitem[Warton and Guttorp(2011)Warton and Guttorp]{wargut:2011}
	Warton, D.~I. and Guttorp, P. (2011).
	\newblock Compositional analysis of overdispersed counts using generalized
	estimating equations.
	\newblock {\em Environmental and Ecological Statistics\/}, {\bf 18}, 427--446.
	
	\bibitem[Xekalaki(2014)Xekalaki]{xekala:2014}
	Xekalaki, E. (2014).
	\newblock On the distribution theory of over-dispersion.
	\newblock {\em Journal of Statistical Distribution and Applications\/}, {\bf
		1}(19).
	
\end{thebibliography}

\newpage
\section*{Figures and Tables}
\newpage
\newpage
\begin{figure}[H]
	\centering
	\includegraphics[scale=0.4]{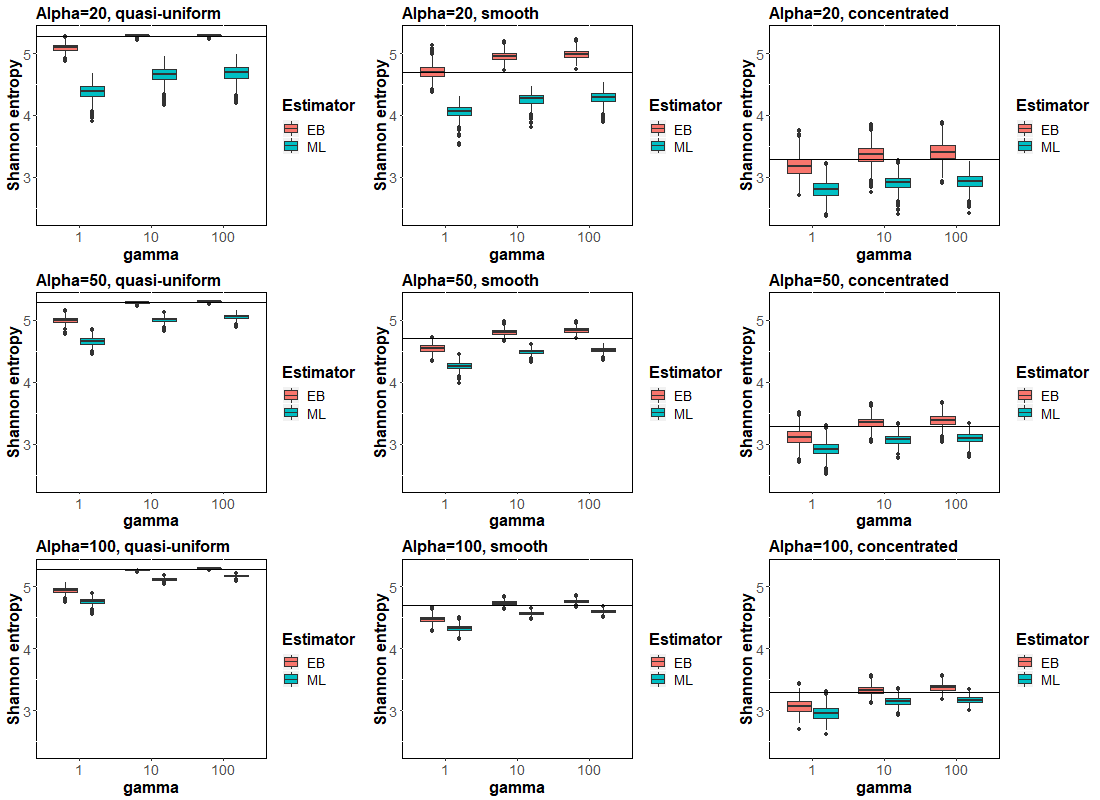}
	\caption{A simulation study on the effects of expected sample size, level of overdispersion and concentration of the population profile on the Shannon entropy $H$. The left column of boxplots is for quasi-uniform profile, the center column for smooth profile and the right column for concentrated profile. The black line in each boxplot represents the respective true value $H^*$ of the index.}\label{fig:boxplot1}
\end{figure}
\newpage
\begin{figure}[H]	
	\centering
	\includegraphics[scale=0.4]{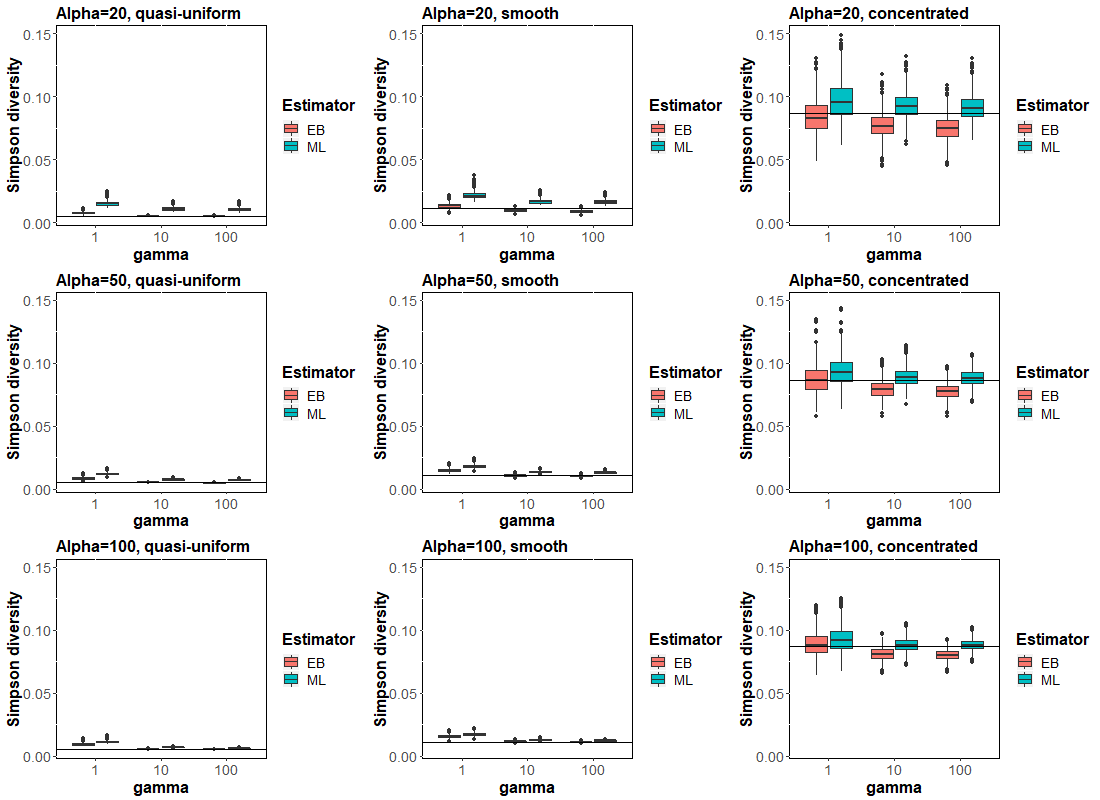}
	\caption{A simulation study on the effects of expected sample size, level of overdispersion and concentration of the population profile on the Simpson diversity $D$. The left column of boxplots is for quasi-uniform profile, the center column for smooth profile and the right column for concentrated profile. The black line in each boxplot represents the respective true value $D^*$ of the index.}\label{fig:boxplot2}
\end{figure}
\newpage
\begin{figure}[H]	
	\centering
	\includegraphics[scale=0.4]{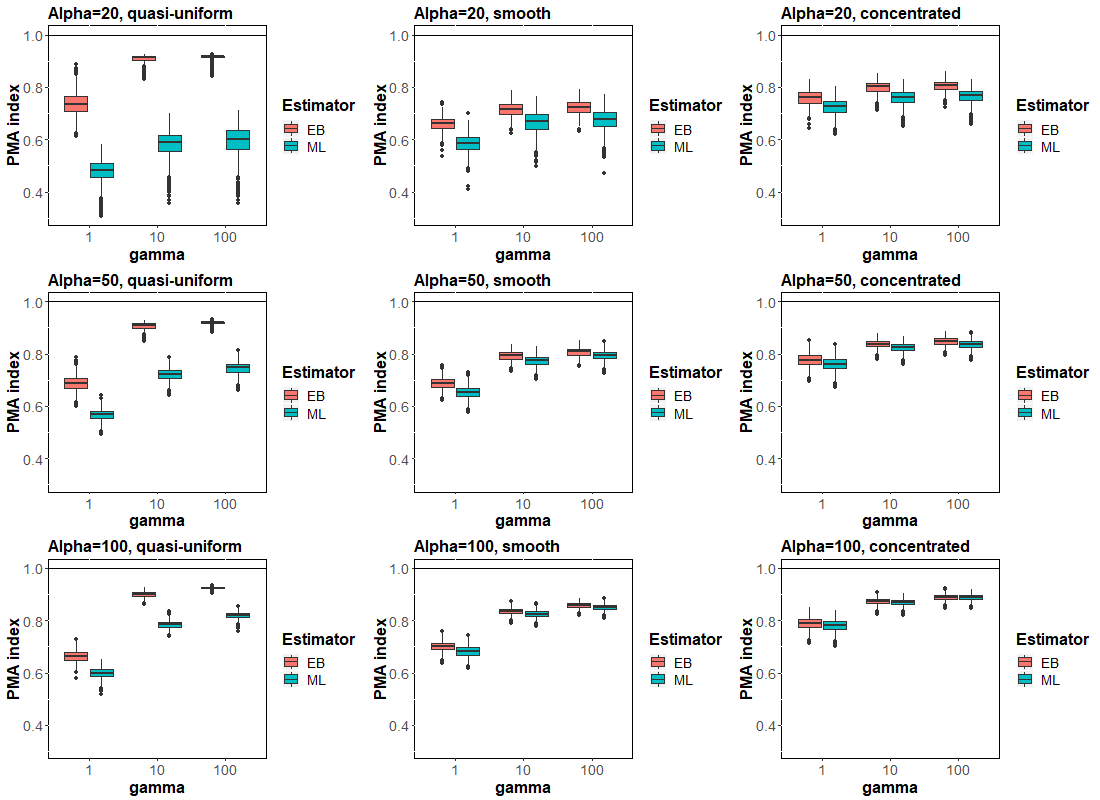}
	\caption{A simulation study on the effects of expected sample size, level of overdispersion and concentration of the population profile on the PMA index $I$. The left column of boxplots is for quasi-uniform profile, the center column for smooth profile and the right column for concentrated profile. The black line in each boxplot represents the respective true value $I^*$ of the index.}\label{fig:boxplot3}
\end{figure}
\newpage
\begin{figure}[H]	
	\centering
	\includegraphics[scale=0.4]{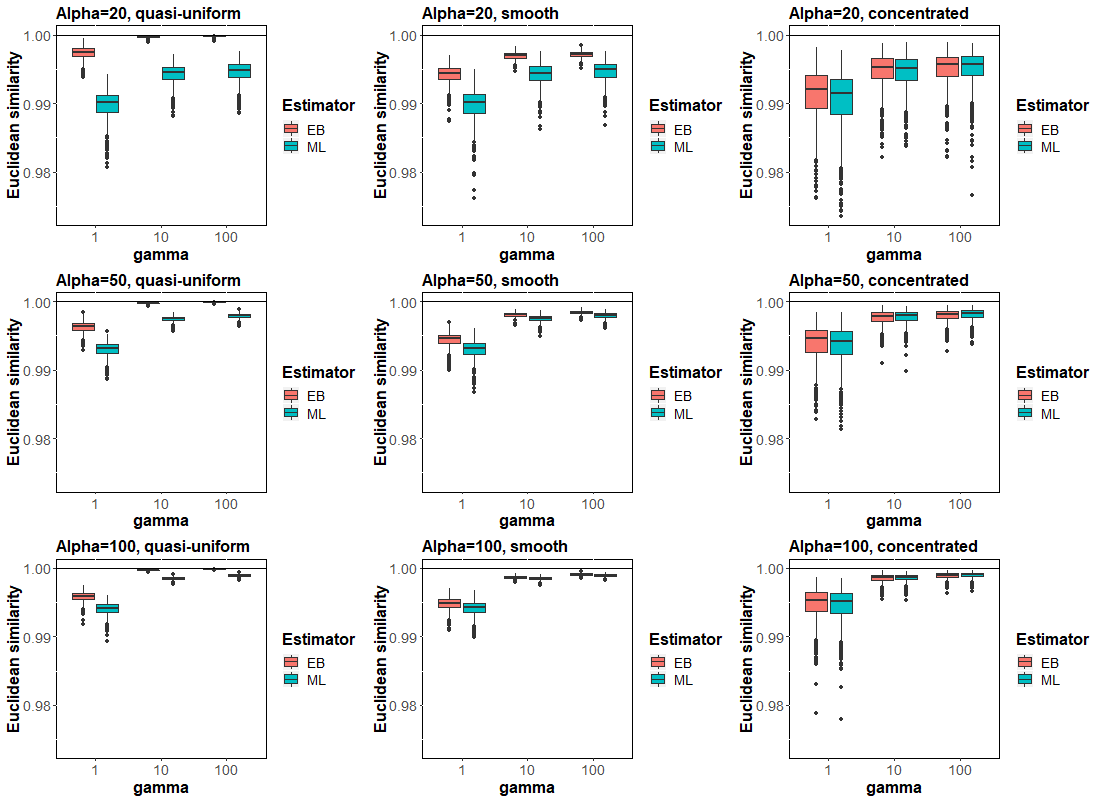}
	\caption{A simulation study on the effects of expected sample size, level of overdispersion and concentration of the population profile on the Euclidian similarity $E$. The left column of boxplots is for quasi-uniform profile, the center column for smooth profile and the right column for concentrated profile. The black line in each boxplot represents the respective true value $E^*$ of the index.}\label{fig:boxplot4}
\end{figure}
\newpage
\begin{table}[H]\footnotesize
	\centering
	\caption{Shannon entropy $H$, sampling mean, standard deviation (SD), bias, and root mean squared error (RMSE) of the empirical Bayes (EB) and maximum likelihood (ML) estimates, with respect to the different simulation experiments.}\label{mse1}
	\begin{tabular}{c|c|rrr|rrr|rrr}
		\toprule
		\multicolumn{2}{c|}{\multirow{2}{*}{Parameters}} & \multicolumn{3}{c|}{$\alpha=20$, $\beta=0.1$} & \multicolumn{3}{c|}{$\alpha=50$, $\beta=0.1$ } & \multicolumn{3}{c}{$\alpha=100$, $\beta=0.1$}\\
		\multicolumn{2}{c|}{}                            & $\gamma=1$ & $\gamma=10$ & $\gamma=100$ & $\gamma=1$ & $\gamma=10$ & $\gamma=100$ & $\gamma=1$ & $\gamma=10$ & $\gamma=100$\\
		\midrule
		\multicolumn{11}{c}{Quasi-uniform evenness, true value $H^*=5.280$}\\
		\midrule
		\multirow{4}{*}{EB}
		& Mean & 5.089& 5.287& 5.293& 4.994& 5.280& 5.294& 4.941& 5.270& 5.293\\
		& SD   & 0.068& 0.013& 0.008& 0.056& 0.012& 0.005& 0.048& 0.011& 0.004\\
		& Bias &-0.191& 0.007& 0.013&-0.286& 0.000& 0.014&-0.339&-0.010& 0.013\\
		& RMSE & 0.203& 0.015& 0.016& 0.291& 0.012& 0.016& 0.342& 0.015& 0.014\\ 
		\midrule
		\multirow{4}{*}{ML} 
		& Mean & 4.380& 4.653& 4.682& 4.655& 5.003& 5.048& 4.758& 5.123& 5.169\\
		& SD   & 0.119& 0.127& 0.134& 0.060& 0.045& 0.044& 0.049& 0.022& 0.017\\
		& Bias &-0.900&-0.627&-0.598&-0.625&-0.277&-0.232&-0.522&-0.157&-0.111\\
		& RMSE & 0.908& 0.639& 0.612& 0.628& 0.280& 0.236& 0.524& 0.158& 0.112\\
		\midrule
		\multicolumn{11}{c}{Smooth evenness, true value $H^*=4.699$}\\
		\midrule
		\multirow{4}{*}{EB} 
		& Mean & 4.703& 4.951& 4.979& 4.539& 4.804& 4.832& 4.471& 4.734& 4.768\\
		& SD   & 0.110& 0.081& 0.078& 0.067& 0.047& 0.042& 0.056& 0.033& 0.029\\
		& Bias & 0.004& 0.252& 0.280&-0.160& 0.105& 0.133&-0.228& 0.035& 0.069\\
		& RMSE & 0.110& 0.265& 0.291& 0.174& 0.114& 0.140& 0.235& 0.048& 0.075\\
		\midrule
		\multirow{4}{*}{ML} 
		& Mean & 4.058& 4.256& 4.280& 4.253& 4.487& 4.513& 4.323& 4.568& 4.599\\
		& SD   & 0.102& 0.100& 0.104& 0.062& 0.044& 0.044& 0.053& 0.031& 0.028\\
		& Bias &-0.641&-0.443&-0.419&-0.446&-0.212&-0.186&-0.376&-0.131&-0.100\\
		& RMSE & 0.649& 0.454& 0.432& 0.450& 0.217& 0.191& 0.380& 0.135& 0.104\\
		\midrule
		\multicolumn{11}{c}{Concentrated evenness, true value $H^*=3.291$}\\
		\midrule
		\multirow{4}{*}{EB} 
		& Mean & 3.182& 3.363& 3.406& 3.108& 3.341& 3.381& 3.072& 3.326& 3.369\\
		& SD   & 0.172& 0.157& 0.156& 0.132& 0.099& 0.098& 0.110& 0.075& 0.066\\
		& Bias &-0.109& 0.072& 0.115&-0.182& 0.050& 0.091&-0.218& 0.035& 0.079\\
		& RMSE & 0.204& 0.173& 0.194& 0.225& 0.111& 0.133& 0.244& 0.083& 0.103\\
		\midrule
		\multirow{4}{*}{ML} 
		& Mean & 2.806& 2.906& 2.932& 2.915& 3.066& 3.090& 2.958& 3.144& 3.172\\
		& SD   & 0.144& 0.123& 0.124& 0.120& 0.087& 0.085& 0.104& 0.070& 0.061\\
		& Bias &-0.485&-0.385&-0.359&-0.375&-0.225&-0.200&-0.333&-0.147&-0.119\\
		& RMSE & 0.506& 0.404& 0.380& 0.394& 0.241& 0.218& 0.349& 0.162& 0.134\\
		\bottomrule
	\end{tabular}
\end{table} 
\newpage
\begin{table}[H]\footnotesize
	\centering
	\caption{Simpson diversity $D$, sampling mean, standard deviation (SD), bias, and root mean squared error (RMSE) of the empirical Bayes (EB) and maximum likelihood (ML) estimates, with respect to the different simulation experiments.}\label{mse2}
	\begin{tabular}{c|c|rrr|rrr|rrr}
		\toprule
		\multicolumn{2}{c|}{\multirow{2}{*}{Parameters}} & \multicolumn{3}{c|}{$\alpha=20$, $\beta=0.1$} & \multicolumn{3}{c|}{$\alpha=50$, $\beta=0.1$ } & \multicolumn{3}{c}{$\alpha=100$, $\beta=0.1$}\\
		\multicolumn{2}{c|}{}                            & $\gamma=1$ & $\gamma=10$ & $\gamma=100$ & $\gamma=1$ & $\gamma=10$ & $\gamma=100$ & $\gamma=1$ & $\gamma=10$ & $\gamma=100$\\
		\midrule
		\multicolumn{11}{c}{Quasi-uniform evenness, true value $D^*=0.005$}\\
		\midrule
		\multirow{4}{*}{EB} 
		& Mean & 0.008& 0.005& 0.005&0.009& 0.005& 0.005& 0.009& 0.005& 0.005\\
		& SD   & 0.001&<0.001&<0.001&0.001&<0.001&<0.001& 0.001&<0.001&<0.001\\
		& Bias & 0.003& 0.000& 0.000&0.004& 0.000& 0.000& 0.004& 0.000& 0.000\\
		& RMSE & 0.003&<0.001&<0.001&0.004&<0.001&<0.001& 0.004&<0.001&<0.001\\
		\midrule
		\multirow{4}{*}{ML} 
		&Mean  & 0.015& 0.011& 0.011& 0.012& 0.008& 0.007& 0.011& 0.007& 0.006\\
		& SD   & 0.002& 0.001& 0.001& 0.001&<0.001&<0.001& 0.001&<0.001&<0.001\\
		& Bias & 0.010& 0.006& 0.006& 0.007& 0.003& 0.002& 0.006& 0.002& 0.001\\
		& RMSE & 0.010& 0.006& 0.006& 0.007& 0.003& 0.002& 0.006& 0.002& 0.001\\
		\midrule
		\multicolumn{11}{c}{Smooth evenness, true value $D^*=0.011$}\\
		\midrule
		\multirow{4}{*}{EB} 
		& Mean & 0.014& 0.009& 0.009& 0.015& 0.011& 0.010& 0.016& 0.011& 0.011\\
		& SD   & 0.002& 0.001& 0.001& 0.001& 0.001& 0.001& 0.001&<0.001&<0.001\\
		& Bias & 0.003&-0.002&-0.002& 0.004& 0.000&-0.001& 0.005& 0.000& 0.000\\
		& RMSE & 0.003& 0.002& 0.002& 0.004& 0.001& 0.001& 0.005&<0.001&<0.001\\
		\midrule
		\multirow{4}{*}{ML} 
		& Mean & 0.021& 0.017& 0.016& 0.018& 0.014& 0.013& 0.017& 0.013& 0.012\\
		& SD   & 0.003& 0.002& 0.002& 0.002& 0.001& 0.001& 0.001&<0.001&<0.001\\
		& Bias & 0.010& 0.006& 0.005& 0.007& 0.003& 0.002& 0.006& 0.002& 0.001\\
		& RMSE & 0.011& 0.006& 0.006& 0.007& 0.003& 0.002& 0.006& 0.002& 0.001\\
		\midrule
		\multicolumn{11}{c}{Concentrated evenness, true value $D^*=0.087$}\\
		\midrule
		\multirow{4}{*}{EB} 
		& Mean & 0.084& 0.077& 0.075& 0.087& 0.079& 0.078& 0.089& 0.081& 0.080\\
		& SD   & 0.013& 0.010& 0.010& 0.011& 0.007& 0.006& 0.010& 0.005& 0.004\\
		& Bias &-0.003&-0.010&-0.012& 0.000&-0.007&-0.009& 0.002&-0.006&-0.007\\
		& RMSE & 0.014& 0.014& 0.015& 0.011& 0.010& 0.011& 0.010& 0.008& 0.008\\
		\midrule
		\multirow{4}{*}{ML} 
		& Mean & 0.097& 0.093& 0.091& 0.093& 0.089& 0.088& 0.092& 0.088& 0.088\\
		& SD   & 0.014& 0.010& 0.010& 0.011& 0.007& 0.006& 0.010& 0.005& 0.004\\
		& Bias & 0.010& 0.006& 0.005& 0.007& 0.002& 0.002& 0.005& 0.001& 0.001\\
		& RMSE & 0.017& 0.012& 0.011& 0.013& 0.007& 0.006& 0.011& 0.005& 0.004\\
		\bottomrule
	\end{tabular}
\end{table} 
\newpage
\begin{table}[H]\footnotesize
	\centering
	\caption{PMA index $I$, sampling mean, standard deviation (SD), bias, and root mean squared error (RMSE) of the empirical Bayes (EB) and maximum likelihood (ML) estimates, with respect to the different simulation experiments.}\label{mse3}
	\begin{tabular}{c|c|rrr|rrr|rrr}
		\toprule
		\multicolumn{2}{c|}{\multirow{2}{*}{Parameters}} & \multicolumn{3}{c|}{$\alpha=20$, $\beta=0.1$} & \multicolumn{3}{c|}{$\alpha=50$, $\beta=0.1$ } & \multicolumn{3}{c}{$\alpha=100$, $\beta=0.1$}\\
		\multicolumn{2}{c|}{}                            & $\gamma=1$ & $\gamma=10$ & $\gamma=100$ & $\gamma=1$ & $\gamma=10$ & $\gamma=100$ & $\gamma=1$ & $\gamma=10$ & $\gamma=100$\\
		\midrule
		\multicolumn{11}{c}{Quasi-uniform evenness, true value $I^*=1.000$}\\
		\midrule
		\multirow{4}{*}{EB} 
		& Mean & 0.738& 0.908& 0.914& 0.689& 0.905& 0.919& 0.665& 0.900& 0.924\\
		& SD   & 0.044& 0.015& 0.009& 0.028& 0.013& 0.005& 0.023& 0.012& 0.004\\
		& Bias &-0.262&-0.092&-0.086&-0.311&-0.095&-0.081&-0.335&-0.100&-0.076\\
		& RMSE & 0.266& 0.093& 0.086& 0.313& 0.096& 0.081& 0.336& 0.101& 0.076\\
		\midrule
		\multirow{4}{*}{ML} 
		& Mean & 0.478& 0.581& 0.593& 0.567& 0.720& 0.745& 0.600& 0.784& 0.818\\
		& SD   & 0.044& 0.055& 0.058& 0.022& 0.022& 0.023& 0.020& 0.014& 0.013\\
		& Bias &-0.522&-0.419&-0.407&-0.433&-0.280&-0.255&-0.400&-0.216&-0.182\\
		& RMSE & 0.524& 0.423& 0.411& 0.433& 0.281& 0.256& 0.401& 0.217& 0.182\\
		\midrule
		\multicolumn{11}{c}{Smooth evenness, true value $I^*=1.000$}\\
		\midrule
		\multirow{4}{*}{EB} 
		& Mean & 0.659& 0.716& 0.721& 0.687& 0.792& 0.807& 0.701& 0.834& 0.857\\
		& SD   & 0.025& 0.026& 0.027& 0.022& 0.017& 0.017& 0.020& 0.012& 0.011\\
		& Bias &-0.341&-0.284&-0.279&-0.313&-0.208&-0.193&-0.299&-0.166&-0.143\\
		& RMSE & 0.342& 0.285& 0.280& 0.314& 0.209& 0.194& 0.299& 0.166& 0.144\\
		\midrule
		\multirow{4}{*}{ML} 
		& Mean & 0.585& 0.665& 0.676& 0.653& 0.774& 0.793& 0.682& 0.825& 0.850\\
		& SD   & 0.037& 0.042& 0.042& 0.024& 0.020& 0.020& 0.020& 0.013& 0.012\\
		& Bias &-0.415&-0.335&-0.324&-0.347&-0.226&-0.207&-0.318&-0.175&-0.150\\
		& RMSE & 0.417& 0.338& 0.327& 0.348& 0.227& 0.208& 0.318& 0.176& 0.150\\
		\midrule
		\multicolumn{11}{c}{Concentrated evenness, true value $I^*=1.000$}\\
		\midrule
		\multirow{4}{*}{EB} 
		& Mean & 0.759& 0.800& 0.805& 0.776& 0.838& 0.847& 0.789& 0.872& 0.888\\
		& SD   & 0.029& 0.023& 0.021& 0.025& 0.017& 0.015& 0.023& 0.013& 0.012\\
		& Bias &-0.241&-0.200&-0.195&-0.224&-0.162&-0.153&-0.211&-0.128&-0.112\\
		& RMSE & 0.243& 0.202& 0.196& 0.226& 0.163& 0.154& 0.212& 0.129& 0.113\\
		\midrule
		\multirow{4}{*}{ML} 
		& Mean & 0.723& 0.760& 0.766& 0.761& 0.825& 0.836& 0.782& 0.869& 0.889\\
		& SD   & 0.032& 0.029& 0.028& 0.025& 0.018& 0.017& 0.023& 0.013& 0.012\\
		& Bias &-0.277&-0.240&-0.234&-0.239&-0.175&-0.164&-0.218&-0.131&-0.111\\
		& RMSE & 0.279& 0.242& 0.236& 0.240& 0.176& 0.165& 0.220& 0.131& 0.112\\
		\bottomrule
	\end{tabular}
\end{table} 
\newpage
\begin{table}[H]\footnotesize
	\centering
	\caption{Euclidean similarity $E$, sampling mean, standard deviation (SD), bias, and root mean squared error (RMSE) of the empirical Bayes (EB) and maximum likelihood (ML) estimates, with respect to the different simulation experiments.}\label{mse4}
	\begin{tabular}{c|c|rrr|rrr|rrr}
		\toprule
		\multicolumn{2}{c|}{\multirow{2}{*}{Parameters}} & \multicolumn{3}{c|}{$\alpha=20$, $\beta=0.1$} & \multicolumn{3}{c|}{$\alpha=50$, $\beta=0.1$ } & \multicolumn{3}{c}{$\alpha=100$, $\beta=0.1$}\\
		\multicolumn{2}{c|}{}                            & $\gamma=1$ & $\gamma=10$ & $\gamma=100$ & $\gamma=1$ & $\gamma=10$ & $\gamma=100$ & $\gamma=1$ & $\gamma=10$ & $\gamma=100$\\
		\midrule
		\multicolumn{11}{c}{Quasi-uniform evenness,  true value $E^*=1.000$}\\
		\midrule
		\multirow{4}{*}{EB} 
		& Mean & 0.997& 0.999& 0.999& 0.996& 0.999& 0.999& 0.996& 0.999& 0.999\\
		& SD   & 0.001&<0.001&<0.001& 0.001&<0.001&<0.001& 0.001&<0.001&<0.001\\
		& Bias &-0.003&-0.001&-0.001&-0.004&-0.001&-0.001&-0.004&-0.001&-0.001\\
		& RMSE & 0.003&<0.001&<0.001& 0.004&<0.001&<0.001& 0.004&<0.001&<0.001\\
		\midrule
		\multirow{4}{*}{ML} 
		& Mean & 0.990& 0.994& 0.995& 0.993& 0.997& 0.998& 0.994& 0.998& 0.999\\
		& SD   & 0.002& 0.001& 0.001& 0.001&<0.001&<0.001& 0.001&<0.001&<0.001\\
		& Bias &-0.010&-0.006&-0.005&-0.007&-0.003&-0.002&-0.006&-0.002&-0.001\\
		& RMSE & 0.010& 0.006& 0.006& 0.007& 0.003& 0.002& 0.006& 0.002& 0.001\\
		\midrule
		\multicolumn{11}{c}{Smooth evenness, true value $E^*=1.000$}\\
		\midrule
		\multirow{4}{*}{EB} 
		& Mean & 0.994& 0.997& 0.997& 0.994& 0.998& 0.998& 0.995& 0.999& 0.999\\
		& SD   & 0.001& 0.001&<0.001& 0.001&<0.001&<0.001& 0.001&<0.001&<0.001\\
		& Bias &-0.006&-0.003&-0.003&-0.006&-0.002&-0.002&-0.005&-0.001&-0.001\\
		& RMSE & 0.006& 0.003& 0.003& 0.006& 0.002& 0.002& 0.005& 0.001& 0.001\\
		\midrule
		\multirow{4}{*}{ML} 
		& Mean & 0.990& 0.994& 0.995& 0.993& 0.997& 0.998& 0.994& 0.999& 0.999\\
		& SD   & 0.002& 0.002& 0.002& 0.001&<0.001& 0.001& 0.001&<0.001&<0.001\\
		& Bias &-0.010&-0.006&-0.005&-0.007&-0.003&-0.002&-0.006&-0.001&-0.001\\
		& RMSE & 0.010& 0.006& 0.006& 0.007& 0.003& 0.002& 0.006& 0.002& 0.001\\
		\midrule
		\multicolumn{11}{c}{Concentrated evenness, true value $E^*=1.000$}\\
		\midrule
		\multirow{4}{*}{EB} 
		& Mean & 0.991& 0.995& 0.995& 0.994& 0.998& 0.998& 0.995& 0.999& 0.999\\
		& SD   & 0.004& 0.002& 0.002& 0.002& 0.001& 0.001& 0.002& 0.001&<0.001\\
		& Bias &-0.009&-0.005&-0.005&-0.006&-0.002&-0.002&-0.005&-0.001&-0.001\\
		& RMSE & 0.009& 0.006& 0.005& 0.006& 0.003& 0.002& 0.006& 0.001& 0.001\\
		\midrule
		\multirow{4}{*}{ML} 
		& Mean & 0.991& 0.995& 0.995& 0.994& 0.998& 0.998& 0.995& 0.999& 0.999\\
		& SD   & 0.004& 0.003& 0.002& 0.003& 0.001& 0.001& 0.002& 0.001&<0.001\\
		& Bias &-0.009&-0.005&-0.005&-0.006&-0.002&-0.002&-0.005&-0.001&-0.001\\
		& RMSE & 0.010& 0.006& 0.005& 0.007& 0.003& 0.002& 0.006& 0.001& 0.001\\
		\bottomrule
	\end{tabular}
\end{table}
\begin{sidewaystable}
\begin{table}[H]
	\footnotesize
	\centering
	\caption{Relative efficiency $\text{Eff}_{EB/ML}$ of the empirical Bayes method compared to the maximum likelihood method, with respect to the different simulation scenarios, for each profile (quasi-uniform, smooth, concentrated), and for each indicator ($H$, $D$, $I$, and $E$).}\label{eff}
	\begin{tabular}{c|rrr|rrr|rrr|c}
		\toprule
		\multirow{2}{*}{Parameters}& \multicolumn{3}{c|}{$\alpha=20$, $\beta=0.1$} & \multicolumn{3}{c|}{$\alpha=50$, $\beta=0.1$} & \multicolumn{3}{c|}{$\alpha=100$, $\beta=0.1$}& Total and\\
		& $\gamma=1$ & $\gamma=10$ & $\gamma=100$ & $\gamma=1$ & $\gamma=10$ & $\gamma=100$ & $\gamma=1$ & $\gamma=10$ & $\gamma=100$& partial\\
		\midrule
		\multicolumn{10}{c|}{Shannon entropy $H$}& 2.4\\
		\midrule
		Quasi-uniform& 4.5&42.4&38.0& 2.2&23.1&15.1& 1.5&10.9& 7.9& 3.2\\
		Smooth       & 5.9& 1.7& 1.5& 2.6& 1.9& 1.4& 1.6& 2.8& 1.4& 2.1\\
		Concentrated & 2.5& 2.3& 2.0& 1.7& 2.2& 1.6& 1.4& 2.0& 1.3& 1.9\\
		\midrule
		\multicolumn{10}{c|}{Simpson diversity $D$}& 1.1\\		
		\midrule
		Quasi-uniform& 3.9&37.3&34.4& 1.9&19.8&13.7& 1.4& 9.7& 7.7& 2.7\\
		Smooth       & 3.4& 3.0& 2.3& 1.7& 3.6& 2.4& 1.3& 3.2& 2.4& 2.1\\
		Concentrated & 1.3& 0.8& 0.7& 1.2& 0.7& 0.6& 1.1& 0.7& 0.6& 0.9\\
		\midrule
		\multicolumn{10}{c|}{PMA index $I$}& 1.4\\		
		\midrule
		Quasi-uniform& 2.0& 4.5& 4.8& 1.4& 2.9& 3.1& 1.2& 2.1& 2.4& 1.9\\
		Smooth       & 1.2& 1.2& 1.2& 1.1& 1.1& 1.1& 1.1& 1.1& 1.0& 1.1\\
		Concentrated & 1.1& 1.2& 1.2& 1.1& 1.1& 1.1& 1.0& 1.0& 1.0& 1.1\\
		\midrule
		\multicolumn{10}{c|}{Euclidean similarity $E$}& 1.4\\
		\midrule
		Quasi-uniform& 3.7&20.5&25.0& 1.9& 8.5&11.2& 1.4& 4.5& 6.3& 2.6\\
		Smooth       & 1.8& 1.9& 1.9& 1.3& 1.3& 1.3& 1.1& 1.2& 1.2& 1.5\\
		Concentrated & 1.1& 1.0& 1.0& 1.1& 1.0& 1.0& 1.0& 1.0& 1.0& 1.1\\
		\bottomrule
	\end{tabular}
\end{table}
\end{sidewaystable}
\newpage
\begin{figure}[H]	
	\centering
	\includegraphics[scale=0.35]{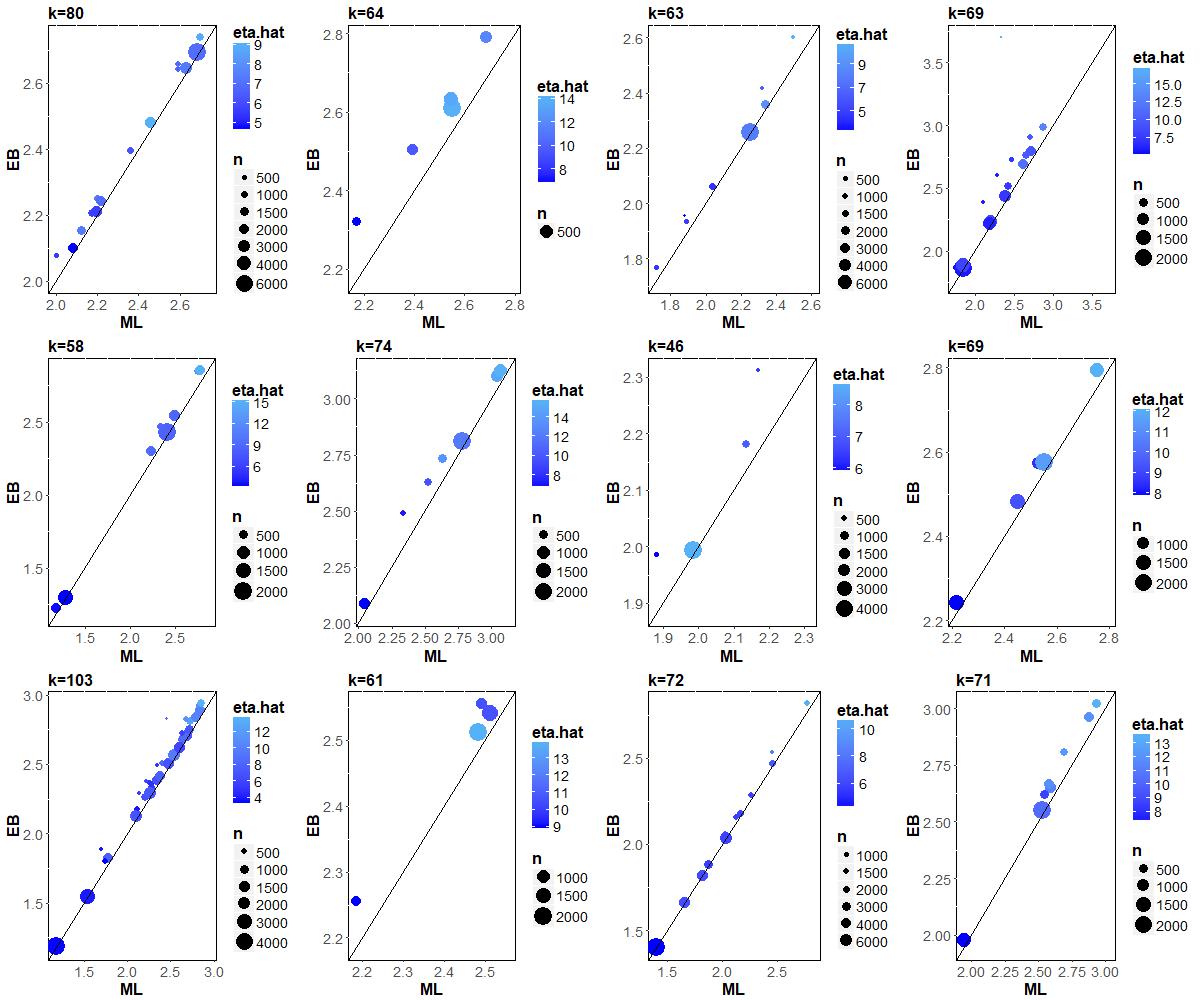}
	\caption{Shannon entropy estimated from benthic macroinvertebrate monitoring data using the maximum likelihood (ML, x-axis) and the empirical Bayes (EB, y-axis) estimators. The 12 plots represent reference site samples from 12 different stream types monitored in Finland. The size of the points is determined by the size of each sample.}\label{fig:app1}
\end{figure}
\newpage
\begin{figure}[H]	
	\centering
	\includegraphics[scale=0.35]{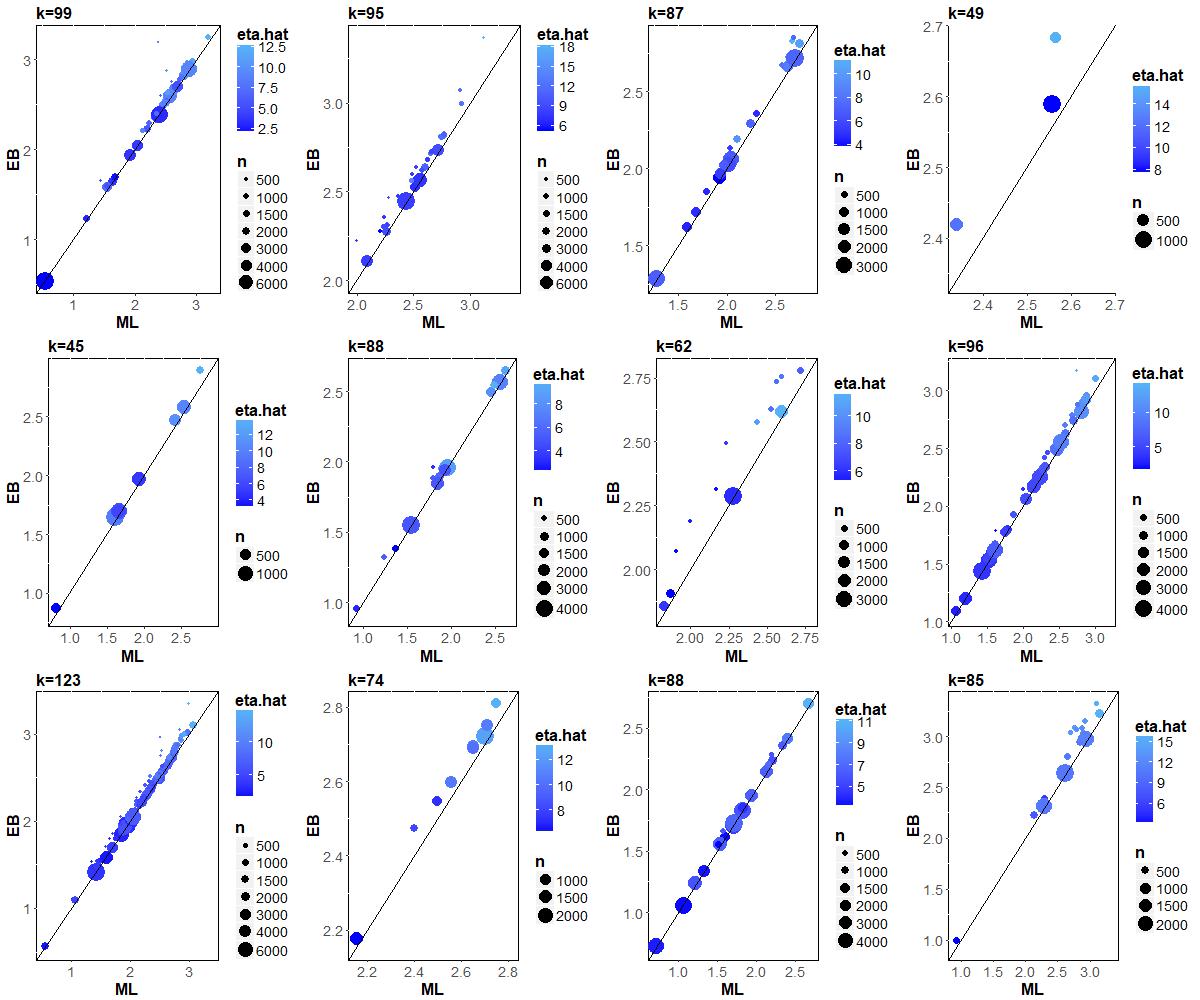}
	\caption{Shannon entropy estimated from benthic macroinvertebrate monitoring data using the maximum likelihood (ML, x-axis) and the empirical Bayes (EB, y-axis) estimators. The 12 plots represent impacted site samples from 12 different stream types monitored in Finland. The size of the points is determined by the size of each sample.}\label{fig:app2}
\end{figure}
\newpage
\begin{figure}[H]
	\centering
	\includegraphics[scale=0.35]{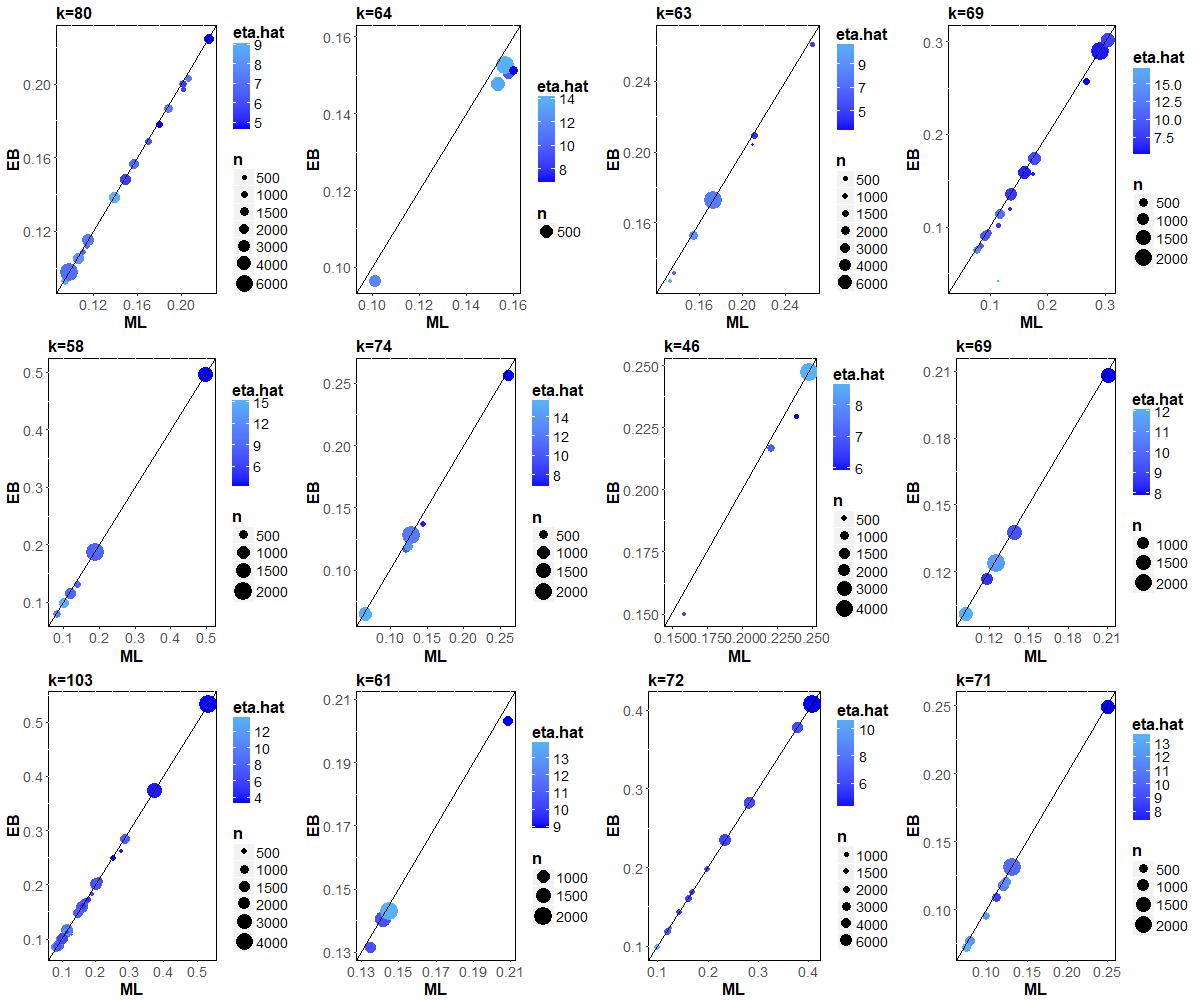}
	\caption{Simpson diversity index estimated from benthic macroinvertebrate monitoring data using the maximum likelihood (ML, x-axis) and the empirical Bayes (EB, y-axis) estimators. The 12 plots represent reference site samples from 12 different stream types monitored in Finland. The size of the points is determined by the size of each sample.}\label{fig:app3}
\end{figure}
\newpage
\begin{figure}[H]
	\centering
	\includegraphics[scale=0.35]{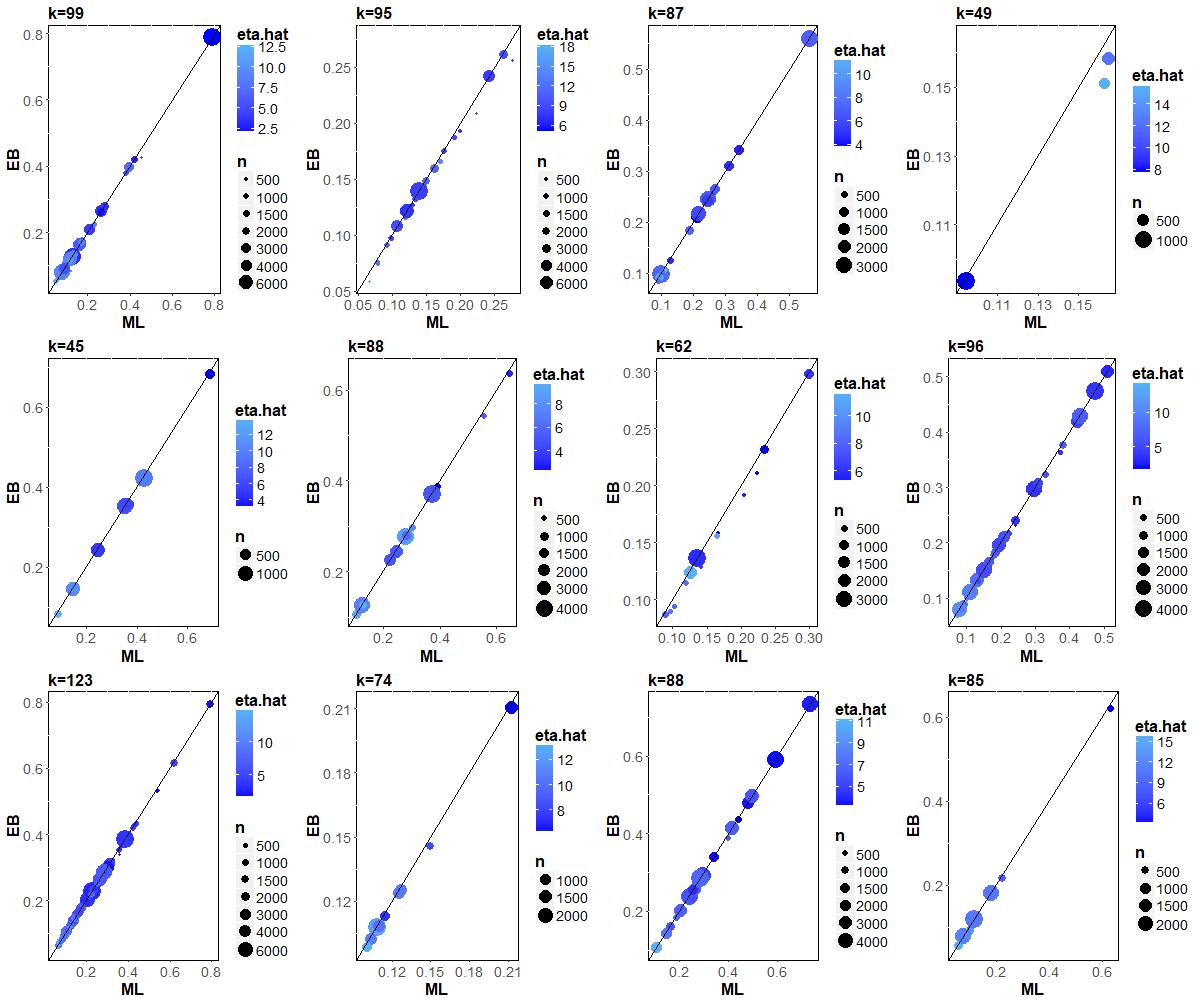}
	\caption{Simpson diversity index estimated from benthic macroinvertebrate monitoring data using the maximum likelihood (ML, x-axis) and the empirical Bayes (EB, y-axis) estimators. The 12 plots represent impacted site samples from 12 different stream types monitored in Finland. The size of the points is determined by the size of each sample.}\label{fig:app4}
\end{figure}
\newpage
\section*{Appendix A: Sampling quantiles by indicator from the simulation study}
\begin{table}[H]\footnotesize
	\centering
	\caption{Shannon entropy, sampling quantiles of the empirical Bayes (EB) and maximum likelihood (ML) estimates with respect to the different simulation scenarios.}\label{tab1}
	\begin{tabular}{c|c|rrr|rrr|rrr}
		\toprule
		\multicolumn{2}{c|}{\multirow{2}{*}{Parameters}} & \multicolumn{3}{c|}{$\alpha=20$, $\beta=0.1$} & \multicolumn{3}{c|}{$\alpha=50$, $\beta=0.1$ } & \multicolumn{3}{c}{$\alpha=100$, $\beta=0.1$}\\
		\multicolumn{2}{c|}{}                            & $\gamma=1$ & $\gamma=10$ & $\gamma=100$ & $\gamma=1$ & $\gamma=10$ & $\gamma=100$ & $\gamma=1$ & $\gamma=10$ & $\gamma=100$\\
		\midrule
		\multicolumn{11}{c}{Quasi-uniform evenness, true value $H^*=5.280$}\\
		\midrule
		\multirow{5}{*}{EB} 
		&Min      &	4.865&	5.213&	5.225&	4.762&	5.222&	5.258&	4.746&	5.228&	5.272\\
		&1st Q.   &	5.048&	5.281&	5.291&	4.958&	5.273&	5.293&	4.911&	5.263&	5.291\\
		&Median   &	5.090&	5.291&	5.298&	4.994&	5.281&	5.296&	4.942&	5.271&	5.294\\
		&3rd Q.   &	5.134&	5.297&	5.298&	5.029&	5.289&	5.298&	4.975&	5.277&	5.296\\
		&Max      &	5.267&	5.298&	5.298&	5.159&	5.298&	5.298&	5.068&	5.296&	5.298\\
		\midrule
		\multirow{5}{*}{ML} 
		&Min      &	3.905&	4.163&	4.191&	4.451&	4.817&	4.874&	4.548&	5.041&	5.091\\
		&1st Q.   &	4.310&	4.577&	4.599&	4.615&	4.976&	5.020&	4.729&	5.110&	5.160\\
		&Median   &	4.390&	4.668&	4.690&	4.658&	5.006&	5.051&	4.760&	5.124&	5.171\\
		&3rd Q.   &	4.463&	4.738&	4.781&	4.697&	5.035&	5.077&	4.792&	5.139&	5.181\\
		&Max      &	4.671&	4.952&	4.987&	4.846&	5.125&	5.156&	4.898&	5.189&	5.214\\
		\midrule
		\multicolumn{11}{c}{Smooth evenness, true value $H^*=4.699$}\\
		\midrule
		\multirow{5}{*}{EB} 
		&Min      &	4.376&	4.727&	4.739&	4.329&	4.663&	4.703&	4.286&	4.630&	4.669\\
		&1st Q.   &	4.631&	4.896&	4.928&	4.494&	4.772&	4.802&	4.433&	4.711&	4.750\\
		&Median   &	4.698&	4.948&	4.980&	4.537&	4.804&	4.832&	4.470&	4.735&	4.769\\
		&3rd Q.   &	4.771&	5.004&	5.033&	4.584&	4.833&	4.861&	4.511&	4.755&	4.786\\
		&Max      &	5.130&	5.200&	5.227&	4.721&	4.972&	4.980&	4.667&	4.841&	4.855\\
		\midrule
		\multirow{5}{*}{ML} 
		&Min      &	3.519&	3.813&	3.890&	3.972&	4.323&	4.342&	4.149&	4.468&	4.503\\
		&1st Q.   &	3.997&	4.198&	4.221&	4.213&	4.460&	4.487&	4.287&	4.548&	4.581\\
		&Median   &	4.066&	4.267&	4.293&	4.255&	4.489&	4.515&	4.323&	4.568&	4.600\\
		&3rd Q.   &	4.127&	4.324&	4.350&	4.297&	4.517&	4.544&	4.361&	4.588&	4.619\\
		&Max      &	4.308&	4.472&	4.530&	4.443&	4.612&	4.618&	4.509&	4.655&	4.678\\
		\midrule
		\multicolumn{11}{c}{Concentrated evenness, true value $H^*=3.291$}\\
		\midrule
		\multirow{5}{*}{EB} 
		&Min      &	2.707&	2.752&	2.901&	2.707&	3.021&	3.026&	2.696&	3.115&	3.188\\
		&1st Q.   &	3.067&	3.262&	3.302&	3.020&	3.277&	3.318&	2.994&	3.277&	3.324\\
		&Median   &	3.183&	3.365&	3.400&	3.102&	3.340&	3.381&	3.075&	3.328&	3.368\\
		&3rd Q.   &	3.292&	3.458&	3.516&	3.198&	3.403&	3.446&	3.146&	3.374&	3.413\\
		&Max      &	3.765&	3.857&	3.884&	3.515&	3.648&	3.673&	3.434&	3.565&	3.573\\		
		\midrule
		\multirow{5}{*}{ML} 
		&Min      &	2.369&	2.394&	2.411&	2.507&	2.772&	2.781&	2.608&	2.915&	3.008\\
		&1st Q.   &	2.709&	2.834&	2.855&	2.840&	3.005&	3.033&	2.881&	3.098&	3.131\\
		&Median   &	2.810&	2.913&	2.932&	2.912&	3.067&	3.092&	2.960&	3.144&	3.171\\
		&3rd Q.   &	2.908&	2.986&	3.019&	2.999&	3.124&	3.149&	3.028&	3.189&	3.211\\
		&Max      &	3.217&	3.271&	3.258&	3.299&	3.340&	3.334&	3.311&	3.363&	3.348\\
		\bottomrule
	\end{tabular}
\end{table} 
\newpage
\begin{table}[H]\footnotesize
	\centering
	\caption{Simpson diversity, sampling quantiles of the empirical Bayes (EB) and maximum likelihood (ML) estimates with respect to the different simulation scenarios.}\label{tab2}
	\begin{tabular}{c|c|rrr|rrr|rrr}
		\toprule
		\multicolumn{2}{c|}{\multirow{2}{*}{Parameters}} & \multicolumn{3}{c|}{$\alpha=20$, $\beta=0.1$} & \multicolumn{3}{c|}{$\alpha=50$, $\beta=0.1$ } & \multicolumn{3}{c}{$\alpha=100$, $\beta=0.1$}\\
		\multicolumn{2}{c|}{}                            & $\gamma=1$ & $\gamma=10$ & $\gamma=100$ & $\gamma=1$ & $\gamma=10$ & $\gamma=100$ & $\gamma=1$ & $\gamma=10$ & $\gamma=100$\\
		\midrule
		\multicolumn{11}{c}{Quasi-uniform evenness, true value $D^*=0.005$}\\
		\midrule
		\multirow{5}{*}{EB} 
		&Min      &	0.005&	0.005&	0.005&	0.007&	0.005&	0.005&	0.007&	0.005&	0.005\\
		&1st Q.   &	0.007&	0.005&	0.005&	0.008&	0.005&	0.005&	0.009&	0.005&	0.005\\
		&Median   &	0.008&	0.005&	0.005&	0.009&	0.005&	0.005&	0.009&	0.005&	0.005\\
		&3rd Q.   &	0.008&	0.005&	0.005&	0.009&	0.005&	0.005&	0.010&	0.005&	0.005\\
		&Max      &	0.011&	0.006&	0.006&	0.013&	0.006&	0.005&	0.014&	0.006&	0.005\\
		\midrule
		\multirow{5}{*}{ML} 
		&Min      &	0.011&	0.008&	0.008&	0.009&	0.007&	0.006&	0.009&	0.006&	0.006\\
		&1st Q.   &	0.014&	0.010&	0.010&	0.011&	0.007&	0.007&	0.010&	0.007&	0.006\\
		&Median   &	0.015&	0.011&	0.010&	0.012&	0.008&	0.007&	0.011&	0.007&	0.006\\
		&3rd Q.   &	0.016&	0.012&	0.011&	0.013&	0.008&	0.007&	0.012&	0.007&	0.006\\
		&Max      &	0.025&	0.017&	0.017&	0.016&	0.010&	0.009&	0.016&	0.007&	0.007\\
		\midrule
		\multicolumn{11}{c}{Smooth evenness, true value $D^*=0.011$}\\
		\midrule
		\multirow{5}{*}{EB} 
		&Min      &	0.007&	0.006&	0.006&	0.012&	0.009&	0.009&	0.012&	0.010&	0.010\\
		&1st Q.   &	0.012&	0.009&	0.008&	0.014&	0.010&	0.010&	0.015&	0.011&	0.011\\
		&Median   &	0.014&	0.009&	0.009&	0.015&	0.011&	0.010&	0.016&	0.011&	0.011\\
		&3rd Q.   &	0.015&	0.010&	0.010&	0.016&	0.011&	0.011&	0.016&	0.012&	0.011\\
		&Max      &	0.022&	0.013&	0.013&	0.021&	0.013&	0.012&	0.021&	0.013&	0.012\\
		\midrule
		\multirow{5}{*}{ML} 
		&Min      &	0.016&	0.013&	0.013&	0.014&	0.012&	0.012&	0.013&	0.011&	0.011\\
		&1st Q.   &	0.020&	0.016&	0.015&	0.017&	0.013&	0.013&	0.016&	0.012&	0.012\\
		&Median   &	0.021&	0.017&	0.016&	0.018&	0.014&	0.013&	0.017&	0.013&	0.012\\
		&3rd Q.   &	0.023&	0.018&	0.017&	0.019&	0.014&	0.014&	0.018&	0.013&	0.012\\
		&Max      &	0.038&	0.026&	0.024&	0.025&	0.017&	0.016&	0.022&	0.015&	0.013\\
		\midrule
		\multicolumn{11}{c}{Concentrated evenness, true value $D^*=0.087$}\\
		\midrule
		\multirow{5}{*}{EB} 
		&Min      &	0.048&	0.045&	0.045&	0.058&	0.058&	0.058&	0.064&	0.065&	0.066\\
		&1st Q.   &	0.075&	0.071&	0.068&	0.080&	0.074&	0.073&	0.082&	0.078&	0.077\\
		&Median   &	0.083&	0.077&	0.075&	0.087&	0.079&	0.078&	0.088&	0.081&	0.080\\
		&3rd Q.   &	0.093&	0.083&	0.081&	0.094&	0.084&	0.082&	0.095&	0.085&	0.083\\
		&Max      &	0.130&	0.117&	0.109&	0.135&	0.103&	0.098&	0.120&	0.098&	0.092\\
		\midrule
		\multirow{5}{*}{ML} 
		&Min      &	0.061&	0.062&	0.065&	0.063&	0.067&	0.069&	0.067&	0.072&	0.074\\
		&1st Q.   &	0.086&	0.086&	0.084&	0.085&	0.084&	0.084&	0.086&	0.084&	0.085\\
		&Median   &	0.096&	0.092&	0.090&	0.093&	0.089&	0.088&	0.091&	0.088&	0.088\\
		&3rd Q.   &	0.106&	0.099&	0.098&	0.101&	0.094&	0.092&	0.099&	0.092&	0.091\\
		&Max      &	0.149&	0.132&	0.130&	0.144&	0.114&	0.107&	0.125&	0.105&	0.102\\
		\bottomrule
	\end{tabular}
\end{table} 
\newpage
\begin{table}[H]\footnotesize
	\centering
	\caption{PMA index, sampling quantiles of the empirical Bayes (EB) and maximum likelihood (ML) estimates with respect to the different simulation scenarios.}\label{tab3}
	\begin{tabular}{c|c|rrr|rrr|rrr}
		\toprule
		\multicolumn{2}{c|}{\multirow{2}{*}{Parameters}} & \multicolumn{3}{c|}{$\alpha=20$, $\beta=0.1$} & \multicolumn{3}{c|}{$\alpha=50$, $\beta=0.1$ } & \multicolumn{3}{c}{$\alpha=100$, $\beta=0.1$}\\
		\multicolumn{2}{c|}{}                            & $\gamma=1$ & $\gamma=10$ & $\gamma=100$ & $\gamma=1$ & $\gamma=10$ & $\gamma=100$ & $\gamma=1$ & $\gamma=10$ & $\gamma=100$\\
		\midrule
		\multicolumn{11}{c}{Quasi-uniform evenness, true value $I^*=1$}\\
		\midrule
		\multirow{5}{*}{EB} 
		&Min     &	0.612&	0.830&	0.841&	0.598&	0.848&	0.881&	0.578&	0.863&	0.905\\
		&1st Q.  &	0.709&	0.902&	0.915&	0.670&	0.899&	0.917&	0.649&	0.892&	0.922\\
		&Median  &	0.736&	0.914&	0.917&	0.688&	0.907&	0.919&	0.664&	0.901&	0.924\\
		&3rd Q.  &	0.764&	0.917&	0.918&	0.705&	0.916&	0.921&	0.680&	0.909&	0.926\\
		&Max     &	0.886&	0.924&	0.926&	0.786&	0.927&	0.930&	0.729&	0.928&	0.933\\
		\midrule
		\multirow{5}{*}{ML} 
		&Min     &	0.307&	0.356&	0.357&	0.491&	0.642&	0.661&	0.519&	0.738&	0.757\\
		&1st Q.  &	0.454&	0.554&	0.562&	0.552&	0.706&	0.730&	0.587&	0.774&	0.810\\
		&Median  &	0.483&	0.590&	0.601&	0.568&	0.721&	0.747&	0.600&	0.784&	0.819\\
		&3rd Q.  &	0.508&	0.617&	0.633&	0.582&	0.736&	0.760&	0.615&	0.793&	0.827\\
		&Max     &	0.583&	0.701&	0.712&	0.641&	0.786&	0.813&	0.653&	0.834&	0.855\\
		\midrule
		\multicolumn{11}{c}{Smooth evenness, true value $I^*=1$}\\
		\midrule
		\multirow{5}{*}{EB} 
		&Min     &	0.538&	0.625&	0.631&	0.623&	0.735&	0.753&	0.636&	0.789&	0.818\\
		&1st Q.  &	0.643&	0.698&	0.704&	0.673&	0.781&	0.796&	0.688&	0.827&	0.849\\
		&Median  &	0.661&	0.717&	0.722&	0.687&	0.793&	0.808&	0.701&	0.835&	0.857\\
		&3rd Q.  &	0.676&	0.735&	0.740&	0.702&	0.804&	0.819&	0.715&	0.843&	0.864\\
		&Max     &	0.743&	0.788&	0.790&	0.756&	0.836&	0.853&	0.759&	0.874&	0.885\\
		\midrule
		\multirow{5}{*}{ML} 
		&Min     &	0.409&	0.499&	0.470&	0.574&	0.704&	0.725&	0.617&	0.780&	0.808\\
		&1st Q.  &	0.563&	0.639&	0.650&	0.637&	0.761&	0.781&	0.669&	0.816&	0.843\\
		&Median  &	0.586&	0.668&	0.679&	0.653&	0.775&	0.794&	0.682&	0.825&	0.851\\
		&3rd Q.  &	0.610&	0.696&	0.705&	0.669&	0.788&	0.807&	0.696&	0.834&	0.859\\
		&Max     &	0.701&	0.765&	0.772&	0.730&	0.828&	0.846&	0.742&	0.866&	0.883\\
		\midrule
		\multicolumn{11}{c}{Concentrated evenness, true value $I^*=1$}\\
		\midrule
		\multirow{5}{*}{EB} 
		&Min     &	0.643&	0.710&	0.723&	0.695&	0.780&	0.793&	0.711&	0.824&	0.848\\
		&1st Q.  &	0.740&	0.785&	0.792&	0.758&	0.827&	0.837&	0.774&	0.864&	0.881\\
		&Median  &	0.760&	0.802&	0.807&	0.776&	0.838&	0.847&	0.790&	0.873&	0.889\\
		&3rd Q.  &	0.779&	0.816&	0.820&	0.794&	0.849&	0.858&	0.805&	0.881&	0.896\\
		&Max     &	0.831&	0.854&	0.862&	0.853&	0.877&	0.885&	0.849&	0.909&	0.923\\
		\midrule
		\multirow{5}{*}{ML} 
		&Min     &	0.618&	0.649&	0.657&	0.671&	0.760&	0.774&	0.700&	0.819&	0.846\\
		&1st Q.  &	0.703&	0.742&	0.749&	0.743&	0.813&	0.825&	0.767&	0.861&	0.881\\
		&Median  &	0.725&	0.763&	0.769&	0.761&	0.826&	0.837&	0.783&	0.870&	0.889\\
		&3rd Q.  &	0.746&	0.780&	0.785&	0.780&	0.837&	0.847&	0.798&	0.878&	0.897\\
		&Max     &	0.804&	0.830&	0.829&	0.836&	0.868&	0.882&	0.840&	0.903&	0.919\\
		\bottomrule
	\end{tabular}
\end{table} 
\newpage
\begin{table}[H]\footnotesize
	\centering
	\caption{Euclidean similarity, sampling quantiles of the empirical Bayes (EB) and maximum likelihood (ML) estimates with respect to the different simulation scenarios.}\label{tab4}
	\begin{tabular}{c|c|rrr|rrr|rrr}
		\toprule
		\multicolumn{2}{c|}{\multirow{2}{*}{Parameters}} & \multicolumn{3}{c|}{$\alpha=20$, $\beta=0.1$} & \multicolumn{3}{c|}{$\alpha=50$, $\beta=0.1$ } & \multicolumn{3}{c}{$\alpha=100$, $\beta=0.1$}\\
		\multicolumn{2}{c|}{}                            & $\gamma=1$ & $\gamma=10$ & $\gamma=100$ & $\gamma=1$ & $\gamma=10$ & $\gamma=100$ & $\gamma=1$ & $\gamma=10$ & $\gamma=100$\\
		\midrule
		\multicolumn{11}{c}{Quasi-uniform evenness, true value $E^*=1$}\\
		\midrule
		\multirow{5}{*}{EB} 
		&Min     &	0.994&	0.999&	0.999&	0.993&	0.999&	0.999&	0.992&	0.999& >0.999\\
		&1st Q.  &	0.997& >0.999& >0.999&	0.996& >0.999& >0.999&	0.995& >0.999& >0.999\\
		&Median  &	0.997& >0.999& >0.999&	0.996& >0.999& >0.999&	0.996& >0.999& >0.999\\
		&3rd Q.  &	0.998& >0.999& >0.999&	0.997& >0.999& >0.999&	0.996& >0.999& >0.999\\
		&Max     & >0.999& >0.999& >0.999&	0.998& >0.999& >0.999&	0.997& >0.999& >0.999\\
		\midrule
		\multirow{5}{*}{ML} 
		&Min     &	0.981&	0.988&	0.988&	0.989&	0.996&	0.996&	0.989&	0.998&	0.998\\
		&1st Q.  &	0.989&	0.994&	0.994&	0.992&	0.997&	0.998&	0.994&	0.998&	0.999\\
		&Median  &	0.990&	0.995&	0.995&	0.993&	0.997&	0.998&	0.994&	0.998&	0.999\\
		&3rd Q.  &	0.991&	0.996&	0.996&	0.994&	0.998&	0.998&	0.995&	0.999&	0.999\\
		&Max     &	0.994&	0.998&	0.998&	0.996&	0.998&	0.999&	0.996&	0.999&	0.999\\
		\midrule
		\multicolumn{11}{c}{Smooth evenness, true value $E^*=1$}\\
		\midrule
		\multirow{5}{*}{EB} 
		&Min     &	0.987&	0.995&	0.995&	0.990&	0.996&	0.997&	0.991&	0.998&	0.998\\
		&1st Q.  &	0.993&	0.997&	0.997&	0.994&	0.998&	0.998&	0.994&	0.999&	0.999\\
		&Median  &	0.994&	0.997&	0.997&	0.995&	0.998&	0.998&	0.995&	0.999&	0.999\\
		&3rd Q.  &	0.995&	0.997&	0.997&	0.995&	0.998&	0.999&	0.995&	0.999&	0.999\\
		&Max     &	0.997&	0.998&	0.998&	0.997&	0.999&	0.999&	0.997&	0.999&	0.999\\
		\midrule
		\multirow{5}{*}{ML} 
		&Min     &	0.976&	0.986&	0.987&	0.987&	0.995&	0.996&	0.990&	0.997&	0.998\\
		&1st Q.  &	0.989&	0.993&	0.994&	0.992&	0.997&	0.998&	0.994&	0.998&	0.999\\
		&Median  &	0.990&	0.994&	0.995&	0.993&	0.997&	0.998&	0.994&	0.998&	0.999\\
		&3rd Q.  &	0.991&	0.995&	0.996&	0.994&	0.998&	0.998&	0.995&	0.999&	0.999\\
		&Max     &	0.995&	0.997&	0.998&	0.996&	0.999&	0.999&	0.997&	0.999&	0.999\\
		\midrule
		\multicolumn{11}{c}{Concentrated evenness, true value $E^*=1$}\\
		\midrule
		\multirow{5}{*}{EB} 
		&Min     &	0.976&	0.982&	0.982&	0.983&	0.991&	0.993&	0.979&	0.995&	0.996\\
		&1st Q.  &	0.989&	0.994&	0.994&	0.993&	0.997&	0.997&	0.994&	0.998&	0.999\\
		&Median  &	0.992&	0.995&	0.996&	0.995&	0.998&	0.998&	0.995&	0.999&	0.999\\
		&3rd Q.  &	0.994&	0.997&	0.997&	0.996&	0.998&	0.999&	0.996&	0.999&	0.999\\
		&Max     &	0.998&	0.999&	0.999&	0.998&	0.999&	0.999&	0.999& >0.999& >0.999\\
		\midrule
		\multirow{5}{*}{ML} 
		&Min     &	0.973&	0.984&	0.977&	0.981&	0.990&	0.994&	0.978&	0.995&	0.997\\
		&1st Q.  &	0.988&	0.993&	0.994&	0.992&	0.997&	0.998&	0.993&	0.998&	0.999\\
		&Median  &	0.991&	0.995&	0.996&	0.994&	0.998&	0.998&	0.995&	0.999&	0.999\\
		&3rd Q.  &	0.994&	0.996&	0.997&	0.996&	0.998&	0.999&	0.996&	0.999&	0.999\\
		&Max     &	0.998&	0.999&	0.999&	0.998&	0.999&	0.999&	0.998&	0.999& >0.999\\
		\bottomrule
	\end{tabular}
\end{table} 
\newpage
\section*{Appendix B: Efficiency and relative efficiency}
Let us denote by $J$ any indices considered in the present work, with $J$=$H$, $D$, $I$, or $E$. As we consider diversity and similarity indicators, $J$ is a specific function of the evenness parameter $\underline{\pi}$, that is
$$
J=J(\underline{\pi}).
$$
Further, let us consider the simulation setting described in Section 4 and the estimate $\hat{J}_i$ of $J$ obtained in correspondence of the sample $S_i$, with $i=1,...,m$. In general, the value $\hat{J}_i$ depends on the estimated profile $\hat{\underline{\pi}}_i$ (computed using EB or ML) but also on the values of the parameters $\alpha$, $\beta$, $\gamma$, and $\underline{\pi}^*$ specified to simulate the data $\underline{x}_i$ of the sample $S_i$. Hence, with obvious notation we can represent $\hat{J}_i$ as
$$
\hat{J}_i=J(\hat{\underline{\pi}}_i;\underline{\phi},\underline{\pi}^*),
$$
where $\underline{\phi}=(\alpha, \beta,\gamma)$. In this setting, $\hat{\underline{\pi}}_i$ may denote $\hat{\underline{\pi}}_{EB}$ or $\hat{\underline{\pi}}_{ML}$ in correspondence of the sample $S_i$.\\
To evaluate the efficiency of the estimation methods, EB and ML respectively, we consider the classic root mean squared error
$$
RMSE(J)=\sqrt{\sum_{i=1}^{m} [ \hat{J}_i - J^*]^2},
$$
where $J^*=J(\underline{\pi}^*)$ denotes the value of the index $J$ with respect to the true profile $\underline{\pi}^*$.\\ 
Let us denote by $\Phi$ the set of the simulation parameters $\phi$ and by $\Pi$ the set of the profiles $\underline{\pi}^*$ considered in our simulation study. Then, to specify the dependence on each scenario $\underline{\phi} \in \Phi$ and profile $\underline{\pi}^* \in \Pi$, the RMSE can be rewritten as follows
$$
RMSE(J;\underline{\phi},\underline{\pi}^*)=\sqrt{\sum_{i=1}^{m} [ J(\hat{\underline{\pi}}_i;\underline{\phi},\underline{\pi}^*) - J(\underline{\pi}^*) ]^2}.
$$
In analogy with regression fitting \citep{dodge:2008}, we want to evaluate the efficiency of each estimation method (partially) for each profile $\underline{\pi}^*$, and (totally) for the indicator $J$. Therefore, we introduce two specific measures: the root (partial) mean squared error 
$$
RMSE(J;\underline{\pi}^*)=\sqrt{\sum_{\underline{\phi}\in\Phi} \sum_{i=1}^{m} [ J(\hat{\underline{\pi}}_i;\underline{\phi},\underline{\pi}^*) - J(\underline{\pi}^*) ]^2}, 
$$
and the root (total) mean squared error
$$
RMSE(J)=\sqrt{\sum_{\underline{\pi}^*\in\Pi} \sum_{\underline{\phi}\in\Phi} \sum_{i=1}^{m} [ J(\hat{\underline{\pi}}_i;\underline{\phi},\underline{\pi}^*) - J(\underline{\pi}^*) ]^2},
$$
respectively. Obviously, these measures can be computed using both the EB and ML methods.\\
In the present work, as measure of relative efficiency, we consider the ratio between the inverse of the empirical Bayes root mean squared error and the inverse of the maximum likelihood root mean squared error. Therefore, for each indicator $J$, we define specific (for each scenario $\phi$ and profile $\underline{\pi}^*$), partial (for each profile $\underline{\pi}^*$), and total measures of relative efficiency as follows
$$
\text{Eff}_{EB/ML}(J;\underline{\phi},\underline{\pi}^*)=\dfrac{RMSE_{ML}(J;\underline{\phi},\underline{\pi}^*)}{RMSE_{EB}(J;\underline{\phi},\underline{\pi}^*)},
$$
$$
\text{Eff}_{EB/ML}(J;\underline{\pi}^*)=\dfrac{RMSE_{ML}(J;\underline{\pi}^*)}{RMSE_{EB}(J;\underline{\pi}^*)},
$$
and
$$
\text{Eff}_{EB/ML}(J)=\dfrac{RMSE_{ML}(J)}{RMSE_{EB}(J)},
$$
with obvious notation.
\end{document}